\documentclass[11pt]{scrartcl}
\usepackage[disable]{todonotes}

\usepackage[utf8]{inputenc} 
\usepackage{mathtools, amsfonts, amssymb, breqn,bm}
\usepackage{amsthm}

\usepackage[british]{babel}
\usepackage{graphicx, hyperref, tabularx, abstract, makecell,ragged2e}

\usepackage[format=plain]{caption}
\DeclareCaptionFormat{cont}{#1 (cont.)#2#3\par}

\usepackage[outdir=./]{epstopdf}   

\usepackage{cellspace, hhline} 

\usepackage{subfig}
\usepackage{xcolor-solarized}
\usepackage{todonotes}
\usepackage{overpic,cleveref}

\usepackage{enumitem}
\setdescription{itemsep=5pt,parsep=0pt,leftmargin=1.55cm,font=\normalfont}

\usepackage[]{geometry}
\newcommand{\R}{\ensuremath{\mathbb{R}}}
\newcommand{\C}{\ensuremath{\mathbb{C}}}

\newcommand{\dif}{{\operatorname{d}}}

\usepackage{cellspace, hhline}
\definecolor{changes}{RGB}{0, 0, 0}
\numberwithin{equation}{section}
\numberwithin{figure}{section}
\numberwithin{table}{section}

\numberwithin{equation}{section}
\numberwithin{figure}{section}
\numberwithin{table}{section}

\begin{document}
	\author{L. Eigentler}
	
	\newcommand{\Addresses}{{
			\bigskip
			\footnotesize
			
			L. Eigentler,  Maxwell Institute for Mathematical Sciences, Department of Mathematics, Heriot-Watt University, Edinburgh EH14 4AS, United Kingdom\par\nopagebreak
			\textit{E-mail address}: \texttt{le8@hw.ac.uk}

	}}
	
	\title{Intraspecific competition in models for vegetation patterns: decrease in resilience to aridity and facilitation of species coexistence}
	\date{\vspace{-1.5cm}}
	\maketitle
	\Addresses

\begin{abstract}
	Patterned vegetation is a characteristic feature of many dryland ecosystems. While plant densities on the ecosystem-wide scale are typically low, a spatial self-organisation principle leads to the occurrence of alternating patches of high biomass and patches of bare soil. Nevertheless, intraspecific competition dynamics {\color{changes}other than competition for water over long spatial scales} are commonly ignored in mathematical models for vegetation patterns. In this paper, I address the impact of {\color{changes}local} intraspecific competition on a modelling framework for banded vegetation patterns. Firstly, I show that in the context of a single-species model, neglecting {\color{changes}local} intraspecific competition leads to an overestimation of a patterned ecosystem's resilience to increases in aridity. Secondly, in the context of a multispecies model, I argue that {\color{changes}local} intraspecific competition is a key element in the successful capture of species coexistence in model solutions representing a vegetation pattern. For both models, a detailed bifurcation analysis is presented to analyse the onset, existence and stability of patterns. Besides the {\color{changes}strengths of local intraspecific competition}, also the the difference between two species has a significant impact on the bifurcation structure, providing crucial insights into the complex ecosystem dynamics. Predictions on future ecosystem dynamics presented in this paper, especially on pattern onset and pattern stability, can aid the development of conservation programs.
\end{abstract}

Keywords: periodic travelling waves; wavetrains; pattern formation; spatial self-organization; numerical continuation; competitive exclusion; bifurcation analysis

\section{Introduction}

Approximately 40\% of the Earth's land mass are classified as drylands \cite{Reynolds2007}. The development of an understanding of ecosystem dynamics in water-deprived areas is of considerable socio-economic importance as a similar proportion of the total human population lives in arid and semi-arid climate zones, where agriculture is an integral part of the economy \cite{Dickovick2014}. A characteristic feature of arid ecosystems is vegetation patterns, which form an interface between continuous vegetation cover and full deserts. 

{\color{changes}A mechanism commonly credited with the self-organisation of plants into alternating patches of biomass and bare soil is a positive feedback loop between local growth of vegetation and resource (water) distribution towards areas of high biomass.} Several processes are the cause of such hydrological heterogeneities; for example the formation of biogenic soil crusts on bare ground that inhibit water infiltration into the soil and induce overland water flow, or the creation of soil moisture gradients due to vertically extended root systems in soil types that allow for fast water diffusion \cite{Meron2016}. A common type of pattern is regular stripes that occur on hillslopes parallel to the contours of the terrain \cite{Valentin1999}.  

Ecosystem functioning heavily depends on plant populations as they constitute basal levels of food webs \cite{Meron2019}. Changes to a vegetation pattern's properties, such as wavelength or recovery time from perturbations, can provide early warning signals of desertification processes, a major threat for economies in drylands \cite{Gowda2018, Saco2018}. However, the large spatial and temporal scales associated with the ecohydrological dynamics of vegetation patterns restrict the acquisition of comprehensive high-quality data to specific properties (e.g. wavelength \cite{Deblauwe2012}) and to short time series. As a consequence, mathematical modelling, and in particular continuum approaches using systems of (potentially nonlocal) PDEs, have been established as a powerful tool to disentangle the complex ecosystem dynamics \cite{Meron2019}. {\color{changes} In broad terms, PDE models for patterned vegetation can be separated into two classes: kernel-based models that consist of a single equation describing the nonlocal plant interactions \cite{Lefever1997, Martinez-Garcia2014, Martinez-Garcia2018, Martinez-Garcia2013a}; and ecohydrological systems of two or more PDEs that explicitly account for the plants' interactions with the resource \cite{Gilad2004, Gilad2007, HilleRisLambers2001, Rietkerk2002, Klausmeier1999}. 
	
	A subclass of kernel-based models captures the formation of vegetation patterns purely through nonlocal intraspecific competition among plants within a certain interaction range, whose size is determined by the horizontal extension of the plants' root network  \cite{Martinez-Garcia2013a, Martinez-Garcia2014}. By contrast, ecohydrological models explain the occurrence of spatiotemporal patterns through a scale dependent feedback between short-range facilitation and long-range competition for water \cite{Rietkerk2008}. Thereby, they commonly neglect any local intraspecific competition dynamics other than competition for water, for example the release of autotoxic pathogens into the soil \cite{Mazzoleni2015} or biomass limits in given areas due to the maximum biomasses of single individuals \cite{Gilad2007a}. In particular, the majority of these theoretical frameworks assume that the rate of plant growth is either independent of the plant density or increasing with biomass. Combined with the pattern formation feedback in such models, this can result in solutions with biomass peaks of very high densities (e.g. \cite{Bennett2019}), a mathematically interesting but ecologically potentially irrelevant feature. A notable exception is the Gilad et al. model \cite{Gilad2004, Gilad2007}, in which the rate of plant growth approaches zero as biomass density increases to its maximum value, and becomes negative for higher plant densities. Nevertheless, due to differences in the various modelling frameworks, the precise impact of local intraspecific competition for resources other than water on the ecosystem dynamics has not been previously addressed in the context of ecohydrological models for vegetation patterns.}

%
%

It is a classical result from Lotka-Volterra competition models that the interplay between intraspecific and interspecific competition can facilitate species coexistence in resource-limited ecosystems, provided intraspecific competition among all species is sufficiently stronger than interspecific competition between them (e.g. \cite{Chesson2000}). In the context of patterned vegetation in drylands, coexistence of herbaceous (\textit{grasses}) and woody (\textit{shrubs and trees}) species is commonly observed, despite the species' competition for water \cite{Seghieri1997}. Previous theoretical studies have successfully captured species coexistence in vegetation patterns by making the assumption that only one plant type contributes to the pattern-forming feedback \cite{Nathan2013, Baudena2013}. Such approaches, however, are based on strong assumptions on differences between plant species, such as contrasting functional responses to soil moisture, and may thus not be applicable in a general setting. In a recent paper, I have shown that strong intraspecific competition of a species superior in its colonisation abilities can provide an alternative explanation for species coexistence that does not rely on such species-specific assumptions. I argued that a deeper understanding of the impact of intraspecific competition in spatially extended, resource-limited ecosystems can be a key ingredient in the explanation of species coexistence \cite{Eigentler2020coexistence_pattern}. 

In this paper, I closely investigate the impact of {\color{changes}local intraspecific competition dynamics other than competition for water on solutions of an ecohydrological model for banded vegetation patterns in semi-arid environments. To distinguish these dynamics from long-range intraspecific competition for water, I use the term \textit{local intraspecific competition} to refer to negative density-dependent effects that are unaffected by plants at other space locations.} The paper is split into two major parts. Firstly, I assess the effects of intraspecific competition on pattern onset, existence and stability in the context of a single-species model by comparing results to those obtained for the corresponding model {\color{changes} which only takes into account intraspecific competition for water (Sec. \ref{sec: Multispecies pattern: single-species})}. Secondly, I extend the results presented in \cite{Eigentler2020coexistence_pattern} to provide more insights into how {\color{changes} local} intraspecific competition can enable species coexistence under competition for a sole limiting resource by performing a comprehensive bifurcation analysis of a multispecies model (Sec. \ref{sec: Multispecies pattern: multispecies model}). In \cite{Eigentler2020coexistence_pattern}, I mainly focus on the impact of changes to {\color{changes}local} intraspecific competition strength of either species on the occurrence of coexistence patterns. By contrast, in this paper, I present details on how results relate to earlier modelling studies {\color{changes} that only consider intraspecific competition for water}. In particular, I investigate how the bifurcation structure, especially the onset mechanisms for coexistence patterns, changes under simultaneous and separate variations of {\color{changes}local} intraspecific competition strengths of both species. Moreover, I address how the similarity between two species affects their ability to coexist. This contrasts with the analysis presented in \cite{Eigentler2020coexistence_pattern} which is restricted to grass-tree coexistence, a parameter setting which corresponds to large species difference in the context of this paper.  Finally, in Sec. \ref{sec: Multispecies pattern: discussion}, I provide an interpretation and discussion of my results.

\section{Single-species model} \label{sec: Multispecies pattern: single-species}
\subsection{Model}

Several modelling frameworks to describe the ecohydrological dynamics in vegetation patterns have been proposed over the last two decades (see \cite{Martinez-Garcia2018, Zelnik2013} for reviews). One system that stands out due to its simplicity is the extended Klausmeier model \cite{Klausmeier1999}, a phenomenological reaction-advection-diffusion system which has been the basis for many model extensions (e.g. \cite{Siero2018, Eigentler2018nonlocalKlausmeier, Eigentler2019integrodifference, Eigentler2018impulsiveflat, Consolo2019a, Gandhi2018, Marasco2014}). To investigate the impact of {\color{changes}local intraspecific competition dynamics other than those for water} on the ecosystem dynamics, I adjust the plant growth rate in the Klausmeier model to account for negative effects of crowding. {\color{changes}The resulting model describes the dynamics between the plant density $u(x,t)$ and the water density $w(x,t)$, where the space coordinate $x\in\R$ increases in the uphill direction of the domain and time $t\ge 0$. After as suitable nondimensionalisation \cite{Klausmeier1999, Sherratt2005}\footnote{the nondimensionalisations in \cite{Klausmeier1999, Sherratt2005} do not include $k = \alpha_1\alpha_2^{1/2}\alpha_3^{-1/2}$, where $\alpha_1$, $\alpha_2$, $\alpha_3$ are the strength of the plant species' {\color{changes} local} intraspecific competition, the constant quantifying the plants' enhancement of resource availability and the water's evaporation rate, respectively.}, the model is}

\begin{subequations}\label{eq: Multispecies pattern: single-species model}
	\begin{align}
	\frac{\partial u }{\partial t} &= \overbrace{u^2w\left(1-\frac{u}{k}\right)}^{\text{plant growth}} - \overbrace{Bu}^{\text{plant loss}} + \overbrace{\frac{\partial^2 u}{\partial x^2}}^{\text{plant dispersal}}, \\
	\frac{\partial w }{\partial t} &= \underbrace{A}_{\text{rainfall}} - \underbrace{w}_{\substack{\color{changes}\text{evaporation} \\ \color{changes}\text{and drainage}}} - \underbrace{u^2w}_{\substack{\text{water uptake} \\ \text{by plants}}} + \underbrace{\nu\frac{\partial w}{\partial x}}_{\substack{\text{water flow}\\ \text{downhill}}} +  \underbrace{d\frac{\partial^2w}{\partial x^2}}_{\substack{\text{water} \\ \text{diffusion}}}.
	\end{align}
\end{subequations}
{\color{changes}The only modification to the extended Klausmeier model occurs in the plant growth term. In the extended Klausmeier model, plant growth is proportional to water consumption by plants, modelled by $u^2w$. The nonlinearity arises due to the short-range facilitation by plants and thus is crucial in capturing the formation of spatiotemporal patterns in the model. The term is the product of the consumer density ($u$), the resource density ($w$), and a term that describes the enhancement of resource availability in existing biomass patches ($u$), e.g. due to an increase in soil permeability caused by plants. While water uptake remains unaffected by the model extension, the rate of plant growth in \eqref{eq: Multispecies pattern: single-species model} is not assumed to increase without bound as the plant density increases. Instead it is mediated by a logistic growth-type term, which accounts for local intraspecific competition among the plant species. This type of intraspecific competition may occur due to plant properties, such as maximum standing biomasses of single individuals \cite{Nathan2013} or the release of autotoxic compounds into the soil \cite{Mazzoleni2015}, but does not correspond to intraspecific competition for water; those dynamics are accounted for explicitly through the interactions with the water density.}
{\color{changes}Moreover, in both the extended Klausmeier model and the extension \eqref{eq: Multispecies pattern: single-species model}, water is added to the system at a constant rate representing precipitation, both evaporation/drainage and plant mortality effects occur at constant rates and plant dispersal is modelled through diffusion. Finally, the water transport terms are derived from shallow-water theory, resulting in an advection (if the terrain is sloped) and diffusion term \cite{Gilad2004}. The diffusion of water was not included in the model's original formulation \cite{Klausmeier1999}, but has become a commonly used addition (e.g. \cite{Siteur2014, Zelnik2013}), which leads to the model being referred to as the \textit{extended} Klausmeier model. In principle, the derivation of the flux from shallow-water theory yields a nonlinear diffusion term, but evidence that model outcomes do not significantly depend on the exact functional form has led to the simpler linear term being well-established \cite{Nathan2013}}. The parameters $A$, $k$, $B$, $\nu$ and $d$ are nondimensional parameters that can be interpreted as rainfall volume, strength of {\color{changes}local} intraspecific competition, rate of plant mortality, speed of water flow downhill and the water diffusion coefficient, respectively. Typical parameter estimates (e.g. \cite{Klausmeier1999}) suggest that $\nu \approx 200$ is large compared to other model parameters, as it reflects the difference between the rate of water advection and the rate of plant diffusion. The terrain's slope, however, is not steep itself. {\color{changes}The derivation of water flow using shallow-water theory is only valid as long as water flow occurs as sheet flow and thus \eqref{eq: Multispecies pattern: single-species model} does not apply if the terrain's gradient exceeds a few percent, consistent with topological data from field observations of banded vegetation patterns \cite{Valentin1999}.}

The (extended) Klausmeier model {\color{changes} neglecting local} intraspecific competition can be obtained from \eqref{eq: Multispecies pattern: single-species model} by taking $k\rightarrow \infty$. This limiting case has been the subject of extensive mathematical analyses, in particular on the onset, existence and stability of spatial patterns \cite{Sherratt2013IV}. Onset of patterned solutions in PDE systems usually occurs at either a Hopf bifurcation of a spatially uniform equilibrium or at a homoclinic solution (but see Sec. \ref{sec: Multispecies pattern: onset and existence} for an exception). Typically, onset loci also form the boundaries of the parameter regions in which patterns exist, unless a fold in the solution branch occurs. {\color{changes} The transition from uniform to patterned vegetation due to increases in aridity occurs at a Hopf bifurcation of a spatially uniform equilibrium, while at low rainfall volumes, patterned solutions terminate in a homoclinic solution \cite{Sherratt2013IV}.} The homoclinic solution also provides a lower bound for the pattern existence region, while the upper bound may occur at higher precipitation levels than those of the Hopf bifurcation due to the occurrence of a fold. A powerful tool in the analytical derivation of the patterns' features is the utilisation of the size of the advection parameter $\nu$, which allows for asymptotic approximations valid to leading order in $\nu$ as $\nu \rightarrow \infty$.

The addition of {\color{changes}local} intraspecific competition does not have a qualitative impact on pattern onset, existence and stability in the model but noteworthy quantitative impacts are observed as detailed below. Besides the desert steady state $\bm{{v_s^D}} = (0,A)$, which exists and is stable in the whole parameter space, \eqref{eq: Multispecies pattern: single-species model} admits a pair of vegetated spatially uniform equilibria given by

\begin{align*}
\bm{{v_s^{\pm}}} = \left(\overline{u^{\pm}}, \overline{w^{\pm}}\right) = \left(\frac{A \pm \sqrt{A^2 - 4B\left(B+\frac{A}{k} \right) }}{2\left(B+\frac{A}{k} \right)}, \frac{A}{1+\left( \overline{u^{\pm}}\right)^2} \right),
\end{align*}
which exist provided 

\begin{align*}
A>A_{\min}^G := 2B\left(\frac{1}{k}  + \sqrt{1+\frac{1}{k^2}}\right).
\end{align*}
The lower branch $\bm{{v_s^-}}$ is unstable, while the upper branch $\bm{{v_s^+}}$ is stable to spatially uniform perturbations if $B<2$. Parameter estimates consistently suggest that plant mortality $B$ remains well below this threshold, and thus the case $B\ge 2$ is not considered in the analysis. As is expected, the plant density of the biologically relevant spatially uniform steady state $\bm{{v_s^+}}$ decreases as the strength of {\color{changes}local} intraspecific competition increases (decrease in $k$).

\subsection{Pattern onset, existence \& stability}
Onset of spatial patterns due to a decrease in precipitation $A$ occurs as $\bm{{v_s^+}}$ loses stability to spatially nonuniform perturbations. This is referred to as a Turing-Hopf bifurcation and different methods to analytically calculate an asymptotic approximation of the rainfall threshold exist \cite{Eigentler2018nonlocalKlausmeier}. In this context, this is best performed in travelling wave coordinates; patterned solutions of \eqref{eq: Multispecies pattern: single-species model} are periodic travelling waves, i.e. solutions that are periodic in space and move in the uphill direction of the domain at a constant speed $c \in \R$, and motivate this approach. The transformation into a comoving frame is achieved by setting $z:=x-ct$, $U(z):=u(x,t)$ and $W(z):=w(x,t)$, which yields the travelling wave ODE system

\begin{subequations}\label{eq: Multispecies pattern: single-species model: tw model}
	\begin{align}
	WU^2\left(1-\frac{U}{k}\right) - B U + c\frac{\dif U}{\dif z} +\frac{\dif^2 U}{\dif z^2} &=0, \\
	A-W - WU^2 + (c+\nu) \frac{\dif W}{\dif z} +d \frac{\dif^2 W}{\dif z^2} &=0.
	\end{align}
\end{subequations}

Patterned solutions of the PDE system \eqref{eq: Multispecies pattern: single-species model} correspond to limit cycles of the ODE system \eqref{eq: Multispecies pattern: single-species model: tw model}. In the PDE setting, the patterns' features, such as their existence, would typically be investigated in a one-dimensional parameter space of a chosen control parameter, here the precipitation volume $A$. However, the transformation into travelling wave coordinates introduces an additional parameter, the migration speed $c$. If patterns exist for a given rainfall level in \eqref{eq: Multispecies pattern: single-species model}, then limit cycles with a range of different migration speeds exist in \eqref{eq: Multispecies pattern: single-species model: tw model} for the same precipitation volume. As a consequence, the patterns' features need to be addressed in a two-dimensional parameter space in the travelling wave coordinates, comprised of the chosen PDE bifurcation parameter and the uphill migration speed $c$.

A convenient tool to investigate pattern onset, existence and stability is numerical continuation, but the size of the slope parameter $\nu$ also allows for an analytical derivation of some properties valid to leading order in $\nu$ as $\nu \rightarrow \infty$. A significant challenge of this approach is posed by the dependence of the parameter region in which patterns exist on the slope parameter $\nu$. In particular, the dependence of both $A$ and $c$ on $\nu$ throughout the parameter region covers several orders of magnitude. For the {\color{changes}standard Klausmeier model}, an extensive analysis of these dynamics exists \cite{Sherratt2011, Sherratt2010, Sherratt2013III, Sherratt2013IV, Sherratt2013V}. The focus of this paper is on $c=O_s(1)$ ($x=O_s(y) \Longleftrightarrow x=O(y)$ but not $x=o(y)$) as $\nu \rightarrow \infty$ but the pattern dynamics in \eqref{eq: Multispecies pattern: single-species model} for both small and large migration speeds are expected to be qualitatively similar to those of the model without {\color{changes}local} intraspecific competition.

The rainfall level the Turing-Hopf bifurcation causing pattern onset due to a destabilisation of the spatially uniform equilibrium is $A=O_s(\sqrt{\nu})$ \cite{Sherratt2013V}. An asymptotic approximation of this critical threshold is found by calculating the corresponding Hopf bifurcation in the travelling wave framework and determining the maximum rainfall level on the loci of Hopf bifurcations in the $(A,c)$ plane. The method follows that used for the (extended) Klausmeier model in \cite{Sherratt2013V, Eigentler2018nonlocalKlausmeier}. The rescaling $U=A/BU^\ast$, $W=B^2/AW^\ast$, $z=1/\sqrt{B}z^\ast$, $c=\sqrt{B}c^\ast$, $\Gamma = A^2/(B^{5/2}\nu)$, $\kappa = Bk/A$ and the assumption that $A=O_s(\sqrt{\nu})$ yields

\begin{subequations}\label{eq: Multispecies pattern: single-species model: tw model leading order}
	\begin{align}
	U' &= \widetilde{U},\\
	\widetilde{U}' &= -c\widetilde{U} - WU^2\left(1-\frac{U}{\kappa}\right) + U, \\
	W' &= -\Gamma\left(1-U^2W\right),
	\end{align}
\end{subequations}
valid to leading order in $\nu$ as $\nu \rightarrow \infty$, after dropping the asterisks for brevity. The Hopf locus in the $(A,c)$ parameter plane is calculated through a linear stability analysis. The eigenvalues $\lambda\in\C$ of the Jacobian matrix of \eqref{eq: Multispecies pattern: single-species model: tw model leading order} are assumed to be purely imaginary, i.e. $\lambda = i\omega$, $\omega \in \R$. This allows the Jacobian's characteristic polynomial to be split into its real and imaginary parts and for $\omega$ to be eliminated. The resulting condition implicitly describes the Hopf-locus. Implicit differentiation facilitates the explicit calculation of the rainfall threshold at which the Turing-Hopf bifurcation occurs.

\begin{figure}
	\centering
	\subfloat[$k=10$ (high {\color{changes}local} intraspecific competition)]{\includegraphics[width=0.48\textwidth]{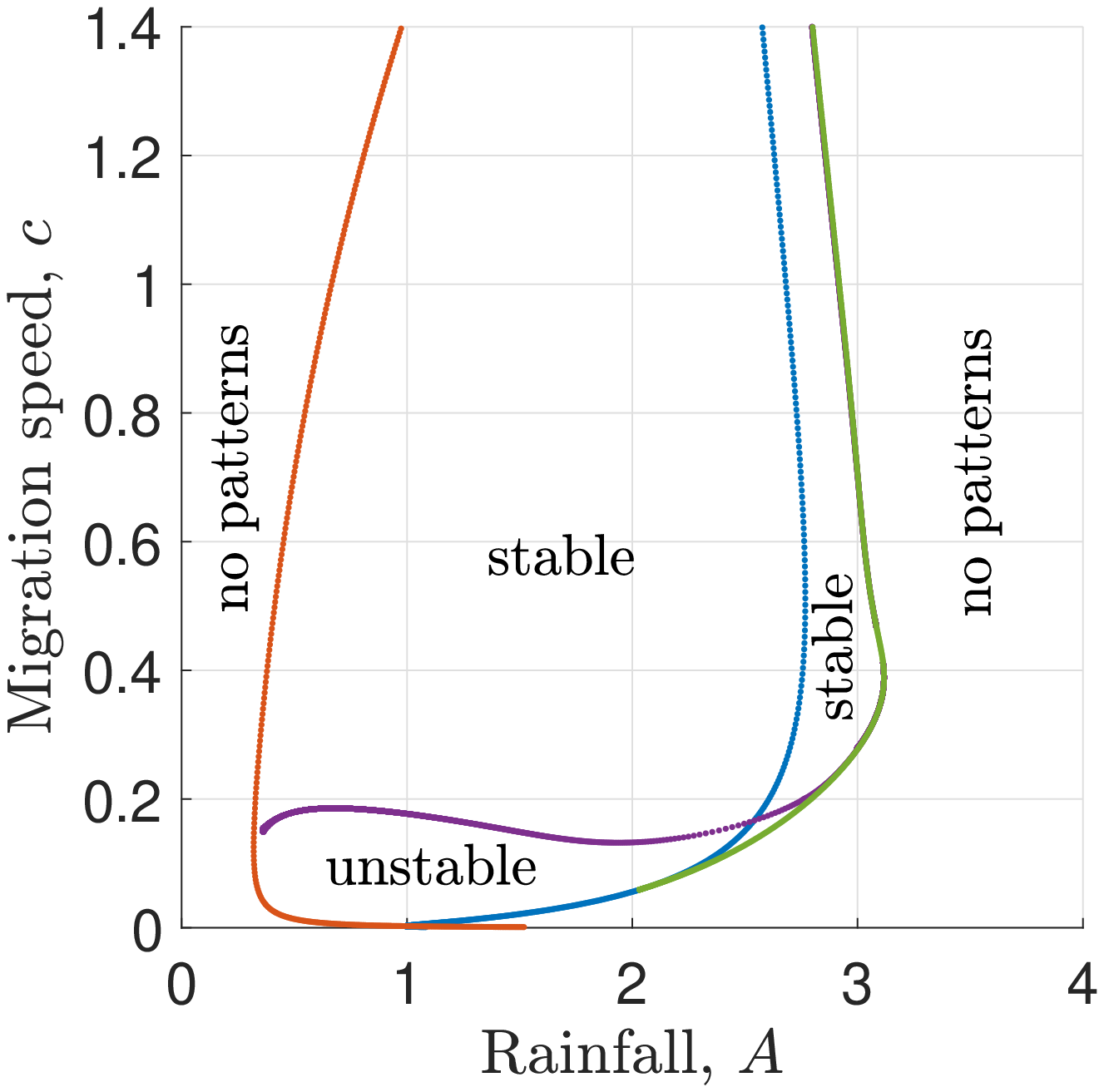}}
	\subfloat[$k=1000$ (low {\color{changes}local} intraspecific competition)]{\includegraphics[width=0.48\textwidth]{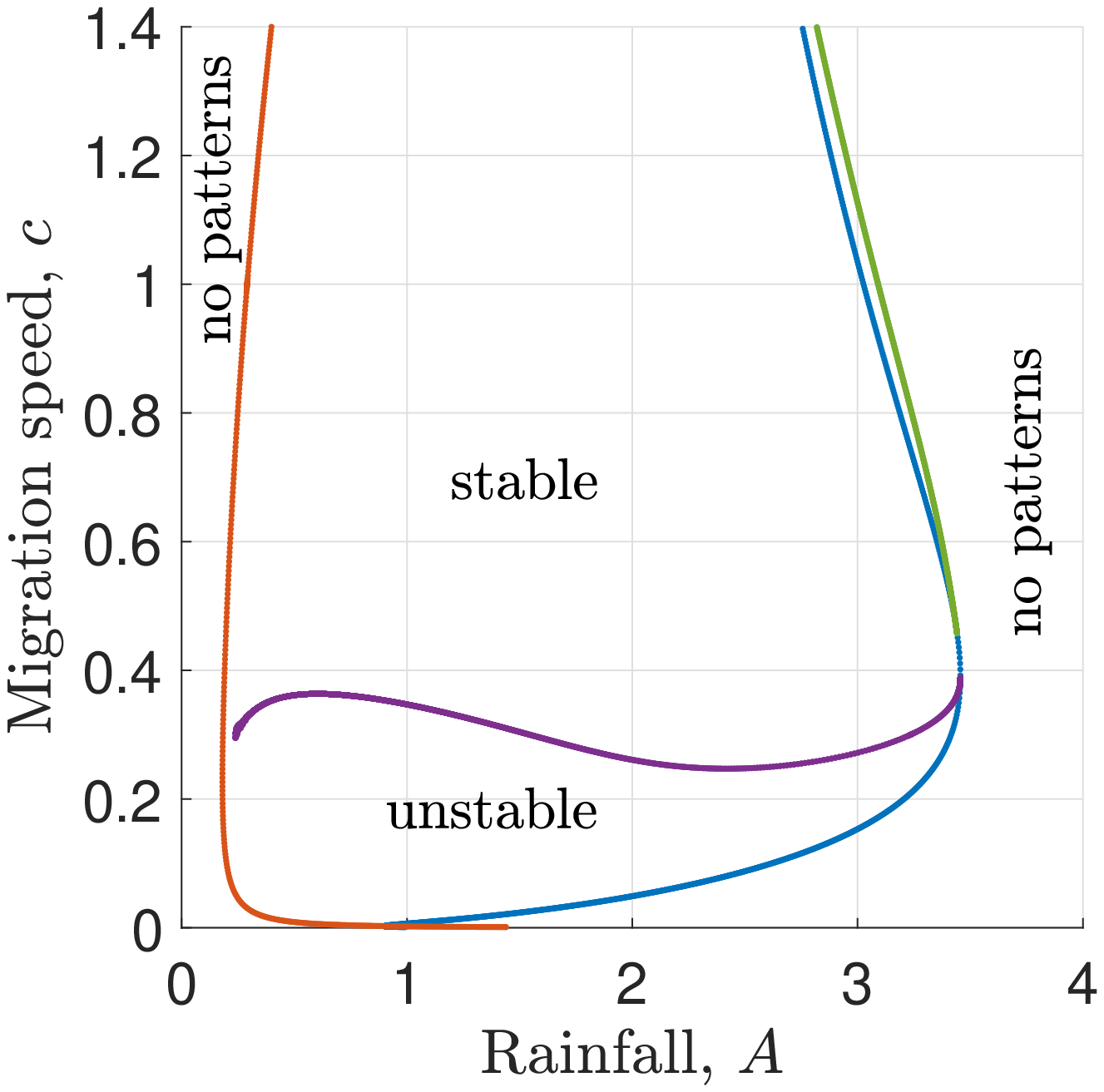}}
	
	\begin{minipage}{0.23\textwidth}
		\vspace*{-6cm}
		\includegraphics[width=\textwidth]{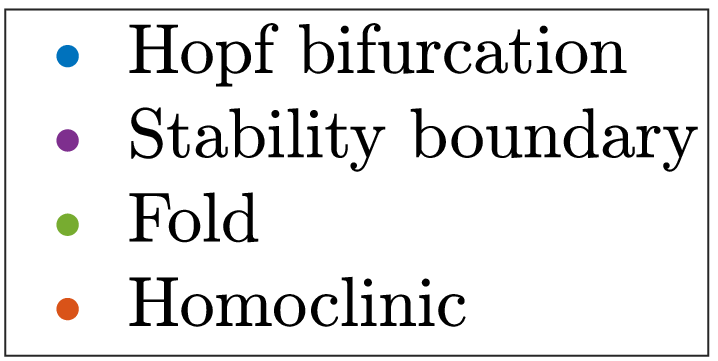}
	\end{minipage}
	\hfill
	\begin{minipage}{0.48\textwidth}
		\subfloat[Relative difference in size of pattern existence rainfall interval\label{fig: Multispecies pattern: single species: pattern existence and stability size diff}]{\includegraphics[width=\textwidth]{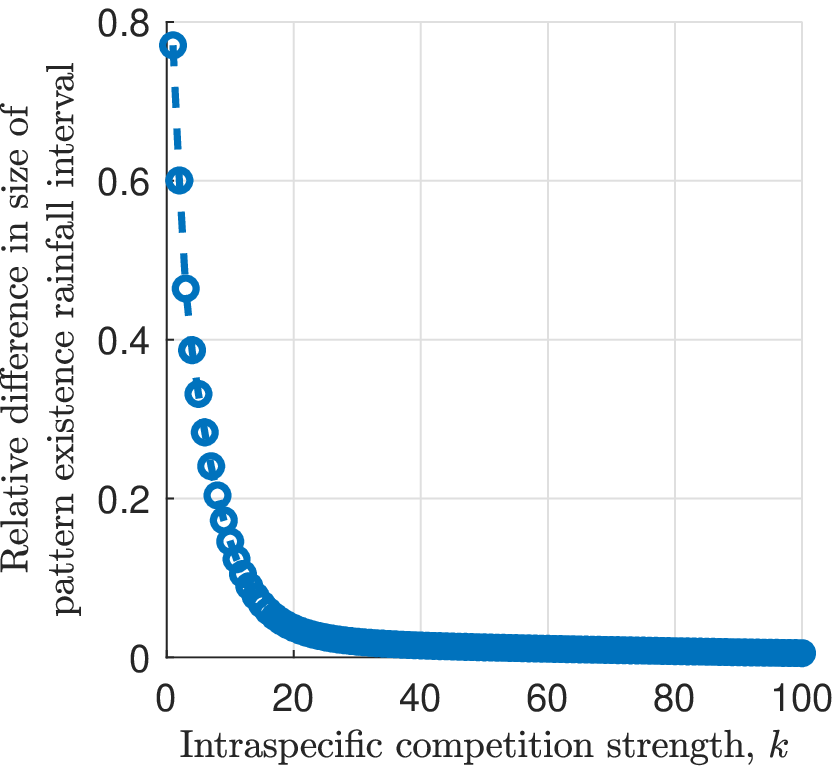}}
	\end{minipage}
	\caption{\textbf{{\color{changes}Local} intraspecific competition stabilises spatially uniform solutions and patterns at lower migration speeds.} Onset, existence and stability parameter regions of patterned solutions of \eqref{eq: Multispecies pattern: single-species model} are shown in the $(A,c)$ parameter plane. Onset at high precipitation values occurs at a Hopf bifurcation, while onset at low values occurs at a homoclinic solution. The existence region of patterns is bounded below by the homoclinic solution and bounded above by either the Hopf bifurcation or a fold in the solution branch, if it exists. Part (a) corresponds to strong {\color{changes}local} intraspecific competition, (b) to weak {\color{changes}local} intraspecific competition. The loci of both the Hopf bifurcation and the fold in the patterned solution branches are shifted to lower precipitation volumes if {\color{changes}local} intraspecific competition is strong, while the homoclinic solution occurs at higher rainfall levels. {\color{changes}Hence, the length of the rainfall interval in which patterns exist decreases with increasing local intraspecific competition. Shown in (c), the relative difference in the size of the pattern existence rainfall interval is given by $(\overline{A}_\infty-\overline{A}_k)/\overline{A}_\infty$, where $\overline{A}_\infty$ and $\overline{A}_k$ are the lengths of the pattern existence rainfall interval in the absence of local intraspecific competition and for local intraspecific competition dynamics with carrying capacity $k$, respectively. Moreover, strong local intraspecific competition stabilises patterns at lower migration speeds.}}\label{fig: Multispecies pattern: single species: pattern existence and stability}
	
\end{figure}

Investigation of this rainfall threshold shows that increases in {\color{changes}local} intraspecific competition shift the Turing-Hopf bifurcation to lower rainfall levels (Fig. \ref{fig: Multispecies pattern: single species: pattern existence and stability}). The stabilisation of the spatially uniform vegetated state is caused by a reduction in plant equilibrium density under strong {\color{changes}local} intraspecific competition which reduces the water requirements of the spatially uniform state.

The subset of the $(A,c)$ parameter plane in which patterned solutions of \eqref{eq: Multispecies pattern: single-species model} exist can be mapped out using numerical continuation. In terms of the PDE control parameter $A$, the pattern existence region is bounded from below by a homoclinic solution. Methods for calculating the location of homoclinic solutions exist \cite{Champneys1996}, but for the analysis presented in this paper it suffices to approximate homoclinic solutions by patterned solutions of large wavelength, say $L=1000$. The upper precipitation bound of the pattern existence parameter region is given by either the Hopf locus or the location of a fold in the solution branch, if such a fold occurs. The impact of {\color{changes}local} intraspecific competition is a reduction in the size of the parameter region in which patterns exist. As discussed above, the Hopf bifurcation occurs at lower rainfall levels if {\color{changes}local} intraspecific competition is strong and the locus of the fold mimics this behaviour. By contrast, the homoclinic solution is located at higher precipitation values if {\color{changes}local} intraspecific competition is strong (Fig. \ref{fig: Multispecies pattern: single species: pattern existence and stability}).

The stability of patterned solutions of \eqref{eq: Multispecies pattern: single-species model} is determined through a calculation of the essential spectrum of the corresponding periodic travelling wave solution in \eqref{eq: Multispecies pattern: single-species model: tw model}. The essential spectrum $\mathcal{S}\subset \C$ of a periodic travelling wave describes the leading order behaviour of perturbations to it. Due to translation invariance of periodic travelling waves, the origin is excluded from the following definition of stability. If $\mathcal{S}$ lies entirely in $\{z\in \C: \Re(z)<0\}$, then the corresponding pattern is stable, otherwise it is unstable. The essential spectrum is calculated through a numerical continuation algorithm by Rademacher et al. \cite{Rademacher2007}, and I refer to \cite{Rademacher2007, Sherratt2012, Sherratt2013a} for full details on the method and to \cite{Eigentler2020savanna_coexistence} for an overview of an implementation to a related system. In particular, the algorithm also facilitates the tracking of stability boundaries, such as that displayed in Fig. \ref{fig: Multispecies pattern: single species: pattern existence and stability} based on a numerical continuation of the spectra. 

An application of this algorithm to \eqref{eq: Multispecies pattern: single-species model} yields that strong {\color{changes}local} intraspecific competition stabilises patterned solutions at slower uphill migration speeds (Fig. \ref{fig: Multispecies pattern: single species: pattern existence and stability}). However, combined with the results on pattern existence discussed above, this also shows that the transition from patterned states to a full desert state occurs at higher rainfall levels if {\color{changes}local} intraspecific competition is strong (Fig. \ref{fig: Multispecies pattern: single species: pattern existence and stability}). {\color{changes}Thus, neglect of intraspecific competition dynamics {\color{changes}other than those for water} in the model causes an overestimation of both the patterns' existence and stability ranges, in particular if a species carrying capacity is small (\cref{fig: Multispecies pattern: single species: pattern existence and stability size diff}).}

\section{Multispecies model} \label{sec: Multispecies pattern: multispecies model}

\subsection{Model}

Species coexistence in dryland ecosystems has previously been addressed in several modelling frameworks. Both Baudena and Rietkerk \cite{Baudena2013} and Nathan et al. \cite{Nathan2013} have successfully explained tree-grass coexistence in patterned form by assuming that only one of the two species induces a pattern-forming feedback loop. The assumption that plant species significantly differ in their functional responses to the environment, however, imposes a restriction on the applicability to a general setting. To overcome this, I have introduced a modelling framework to investigate species coexistence that does not rely on such an assumption in a previous paper \cite{Eigentler2019Multispecies}.

{\color{changes}If intraspecific competition dynamics are restricted to the plant's competition for water}, this model successfully captures species coexistence as long transient states in both a spatially uniform and a vegetation pattern state, provided that species are of similar average fitness \cite{Eigentler2019Multispecies}. Moreover, coexistence is also possible in a \textit{spatially nonuniform savanna state} if there is a balance between the species' local competitiveness and their colonisation abilities \cite{Eigentler2020savanna_coexistence}. {\color{changes}The term \textit{savanna} is ambiguous and a variety of different definitions of savanna ecosystems exist \cite{Taft1997, Scholes1993}. In this paper, \textit{spatially nonuniform savanna} refers to a state that is represented by periodic travelling wave solutions in which both species coexist, their solution profiles are approximately in phase (but see Sec. \ref{sec: Multispecies pattern: phase diff}) and the total plant density oscillates between two nonzero biomass level.} If additionally {\color{changes}local} intraspecific competition dynamics are taken into account, then coexistence is possible in a vegetation pattern state (periodic travelling wave solutions in which the {\color{changes}total} plant density oscillates between a high biomass level and zero), provided {\color{changes}local} intraspecific competition among the superior coloniser is sufficiently large \cite{Eigentler2020coexistence_pattern}. In this paper, I provide more information on the impact of {\color{changes}local} intraspecific competition on the origin and existence of patterned model solutions in which species coexist.

To do so, the model used in the analysis is

\begin{subequations}\label{eq: Multispecies pattern: Model: nondimensional model}
	\begin{align}
	\frac{\partial u_1}{\partial t} &= \overbrace{wu_1\left(u_1 + Hu_2\right) \left(1-\frac{u_1}{k_{1}}\right)}^{\text{plant growth}} - \overbrace{B_1 u_1}^{\substack{\text{plant} \\ \text{mortality}}} + \overbrace{\frac{\partial^2 u_1}{\partial x^2}}^{\text{plant dispersal}}, \label{eq: Multispecies pattern: Model: nondimensional model u1}\\
	\frac{\partial u_2}{\partial t} &=\overbrace{Fwu_2\left(u_1 + Hu_2\right)\left(1-\frac{u_2}{k_2}\right)}^{\text{plant growth}} - \overbrace{B_2 u_2}^{\substack{\text{plant} \\ \text{mortality}}}  +\overbrace{D\frac{\partial^2 u_2}{\partial x^2}}^{\text{plant dispersal}}, \label{eq: Multispecies pattern: Model: nondimensional model u2}\\
	\frac{\partial w}{\partial t} &= \underbrace{A}_{\text{rainfall}}-\underbrace{w}_{\substack{\color{changes}\text{evaporation} \\ \color{changes}\text{and drainage}}} - \underbrace{w\left(u_1+u_2\right)\left(u_1 + Hu_2\right)}_{\text{water uptake by plants}}+ \underbrace{\nu \frac{\partial w}{\partial x}}_{\substack{\text{water flow} \\ \text{ downhill}}} +\underbrace{d \frac{\partial^2 w}{\partial x^2}}_{\substack{\text{water}\\\text{diffusion}}}, \label{eq: Multispecies pattern: Model: nondimensional model water}
	\end{align}
\end{subequations}
after a suitable nondimensionalisation \cite{Eigentler2020coexistence_pattern}. The model is based on the single-species model \eqref{eq: Multispecies pattern: single-species model} presented in Sec. \ref{sec: Multispecies pattern: single-species} and consequently all modelling assumptions are identical to those taken in the single-species model. {\color{changes} In particular, water uptake of species $u_i$ is given by $wu_i(u_1+Hu_2)$ and summing over both species yields the third term in \eqref{eq: Multispecies pattern: Model: nondimensional model water}. In other words, each species not only facilitates its own water consumption (and hence growth) but also that of its competitor. However, the strength of facilitation (for example due to increases soil permeability) differs between species and this is accounted for by the nondimensional constant $H$. As in the single-species model \eqref{eq: Multispecies pattern: single-species model}, plant growth of a species in the absence of local intraspecific competition dynamics is proportional to water consumption of that species. However, to account for local intraspecific competition among species, negative density-dependence is also included in the growth terms. The constants $k_1$ and $k_2$ are the maximum standing biomasses of species $u_1$ and $u_2$, respectively. Note that $u_1$ has no direct competitive impact on $u_2$ and vice versa. Interspecific competition only occurs due to competition for water. The negative density dependence in the growth rates thus strictly correspond to intraspecific competition, for example due to the release of autotoxic pathogens into the soil \cite{Mazzoleni2015}. The parameter $B_1$ of species $u_1$ corresponds to $B$ in the single-species model \eqref{eq: Multispecies pattern: single-species model}, while the additional parameters $F$, $B_2$ and $D$ are all related to the newly introduced species $u_2$ and represent its growth, death rate and dispersal coefficient, respectively.}

Moreover, the single species model \eqref{eq: Multispecies pattern: single-species model} can be obtained from \eqref{eq: Multispecies pattern: Model: nondimensional model} by setting one of the plant densities to zero. In the case of $u_1=0$ this further requires a rescaling. As a consequence, results presented in Sec. \ref{sec: Multispecies pattern: single-species} also hold for the multispecies model \eqref{eq: Multispecies pattern: Model: nondimensional model} in the absence of a competitor species. The introduction of a second species nevertheless has an impact on the single-species states of the system, which is discussed below.

The model {\color{changes}only accounting for intraspecific competition for water is analysed} in \cite{Eigentler2020savanna_coexistence}. It is obtained from \eqref{eq: Multispecies pattern: Model: nondimensional model} by taking the limit $k_1, k_2 \rightarrow \infty$. This limiting behaviour motivates a comparison of results presented in this paper with those in \cite{Eigentler2020savanna_coexistence}, to address what impact the consideration of {\color{changes}local} intraspecific competition dynamics has on the modelling framework. I present results for $k_1=k_2$ to make such a comparison, but also discuss the effects of varying $k_1$ and $k_2$ separately.

The main purpose of this paper is to discuss the impact of {\color{changes}local} intraspecific competition and further develop the understanding of coexistence of herbaceous species and woody species in dryland ecosystems. Due to the symmetry in the model, I assume, without loss of generality, that $u_1$ and $u_2$ represent a grass and tree/shrub species, respectively. {\color{changes}Event though the lack of detailed empirical data does not allow for an accurate parameter estimation, model parameters can be obtained from previous theoretical work (e.g. \cite{Siteur2014, Klausmeier1999}). Moreover, the distinction between a grass and a tree species allows for qualitative assumptions on some model parameters. For example, a plant species' water-to-biomass conversion abilities can be deduced from the time a population requires to attain its steady state density in the absence of any resource scarcity or competition. Grasses reach their equilibrium densities on a much shorter timescale than shrubs and trees, which suggests that they are superior in their ability to convert water into new biomass ($F<1$) \cite{Accatino2010}. Similarly, a species' mortality rate can be inferred from its average lifespan. Typically, grasses have a much shorter lifespan than shrubs and trees which leads to a higher mortality rate in the mathematical model ($B_1>B_2$) \cite{Accatino2010}. The diffusion operators in \eqref{eq: Multispecies pattern: Model: nondimensional model u1} relate the spatial spread of each species with time. Typically, the time from germination to the first dispersal of viable seeds is much longer for shrubs and trees, which suggests a lower diffusion coefficient ($D<1$) \cite{Eigentler2020savanna_coexistence}. Finally, if other parameters are known, the constant describing local facilitation can be deduced from a species' equilibrium density. This is typically higher for shrubs and trees which yields that grasses' facilitative impact per unit biomass is stronger ($H<1$) \cite{Klausmeier1999}. As a consequence of these qualitative assumptions, the grass species $u_1$ is superior in its colonisation abilities and is thus referred to as the \textit{coloniser species} or \textit{pioneer species}.} In the absence of {\color{changes}local} intraspecific competition, species coexistence occurs as a state representing a savanna biome if the inferior coloniser $u_2$ is the \textit{superior local competitor} \cite{Eigentler2020savanna_coexistence}, quantified by the average local fitness difference $B_2-FB_1$ being negative \cite{Eigentler2019Multispecies}. In this paper, I focus on this parameter setting to explore the role of {\color{changes}local} intraspecific competition and species difference in the coexistence of species in vegetation patterns. For the latter, I follow the approach of \cite{Eigentler2019Multispecies} and set

\begin{align} \label{eq: Multispecies pattern: parameter region to compare species}
\begin{split}
B_2 = B_1-\chi(B_1-\widetilde{B_2}), \quad F=1-\chi(1-\widetilde{F}), \quad H=1-\chi(1-\widetilde{H}), \quad D=1-\chi(1-\widetilde{D}),
\end{split} 
\end{align}
where $\widetilde{B_2}$, $\widetilde{F}$, $\widetilde{H}$ and $\widetilde{D}$ are typical parameter estimates for a tree species. Thus, the difference between $u_1$ and $u_2$ is quantified by a single parameter $0\le\chi\le 1$. Note that the {\color{changes}local} intraspecific competition strengths $k_1$ and $k_2$ are not included in this definition as their impact is addressed separately. Unless otherwise stated, I set $B_1=0.45$, $\widetilde{B_2} = 0.004$, $\widetilde{F} =\widetilde{H} = \widetilde{D} =  0.01$, $k_1  = 10$, $k_2=10$, $d=500$ and $\nu = 182.5$ and $\chi=0.9$. The precipitation volume $A$ is the main bifurcation parameter of the system.

\subsection{Stability in spatially uniform model}

\begin{figure}
	\centering
	\includegraphics[width=0.9\textwidth]{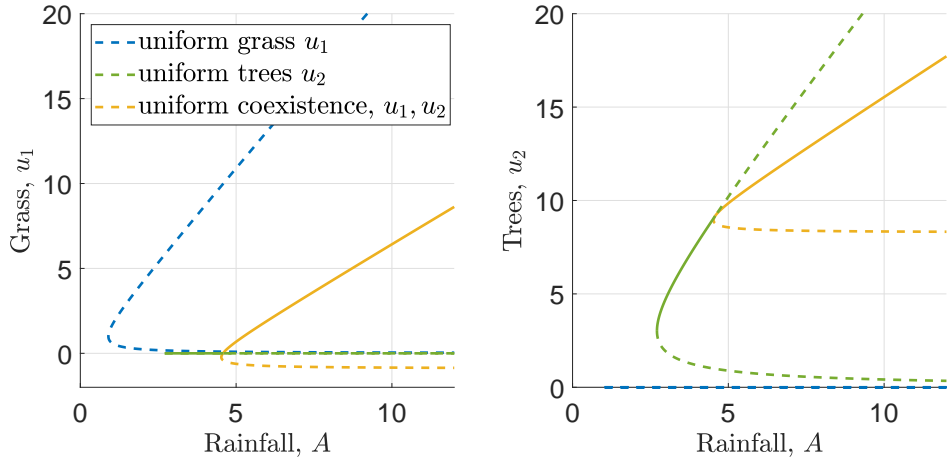}
	\caption{\textbf{Linear stability of spatially uniform equilibria.} The spatially uniform equilibria of \eqref{eq: Multispecies pattern: Model: nondimensional model} and their stability under changes to the precipitation volume $A$ are shown. Solid lines indicate stable states, dashed lines unstable states. For high precipitation values, the coexistence equilibrium $\overline{\bm{{v_m^{c,+}}}}$ is stable because interspecific competition for water is sufficiently lower than intraspecific competition. A decrease in $A$ causes $\overline{\bm{{v_m^{c,+}}}}$ to lose stability to the single-species tree equilibrium $\overline{\bm{{v_m^{t,+}}}}$. For the parameters used in the visualisation the stability change occurs where both equilibria intersect, but this need not be the case. {\color{changes}Also note that at the intersection of equilibria, the coexistence steady state becomes ecologically irrelevant, as one of the plant densities becomes negative. Nevertheless, this steady state can be instructive for mathematical understanding of the dynamics.} The grass equilibrium $\overline{\bm{{v_m^{g,+}}}}$ is unstable for all $A$, because changes in rainfall cannot change which species is of higher local average fitness. Here $k_1=k_2 =1000$ to keep {\color{changes}local} intraspecific competition sufficiently weak. For significantly smaller values of $k_1=k_2$ only the coexistence equilibrium is stable. }\label{fig: Multispecies pattern: spatially uniform bifdiag}
\end{figure}

As for the single-species model \eqref{eq: Multispecies pattern: single-species model}, an understanding of patterned solutions requires knowledge of the system's dynamics in a spatially uniform setting. The system has up to seven spatially uniform equilibria, as visualised in Fig. \ref{fig: Multispecies pattern: spatially uniform bifdiag}. The desert steady state $\overline{\bm{{v_m^d}}} = (0,0,A)$, the pair of single-species grass equilibria $\overline{\bm{{v_m^{g,\pm}}}} = (\overline{u_1^{g,\pm}}, 0,\overline{w^{g,\pm}})$, where $\overline{u_1^{g,\pm}} = \overline{u^\pm}$ and $\overline{w^{g,\pm}} = \overline{w^{\pm}}$ and the latter's existence threshold $A>A_\min^g:=A_{\min}$ are identical with those of the single-species model presented in Sec. \ref{sec: Multispecies pattern: single-species}. Due to the symmetry in the model, \eqref{eq: Multispecies pattern: Model: nondimensional model} also admits a pair of single-species tree equilibria, given by

\begin{align*}
\overline{\bm{{v_m^{t,\pm}}}} := \left(0, \overline{u_2^{t,\pm}}, \overline{w^{t,\pm}}\right) = \left(0, \frac{FHA \pm \sqrt{(FHA)^2 - 4B_2H\left(B_2+\frac{FHA}{k_2} \right) }}{2H\left(B_2+\frac{FHA}{k_2} \right)}, \frac{A}{1+H\left( \overline{u_2^{t,\pm}}\right)^2} \right),
\end{align*}
which exist provided 

\begin{align*}
A>A_{\min}^t := \frac{2B_2}{FH}\left(\frac{1}{k_2}  + \sqrt{H+\frac{1}{k_2^2}}\right).
\end{align*}
Finally, a pair of coexistence spatially uniform steady states $\overline{\bm{{v_m^{c,\pm}}}}:=(\overline{u_1^{c,\pm}},\overline{u_2^{c,\pm}},\overline{w^{c,\pm}})$ exists, provided precipitation is sufficiently large. While it is possible to obtain a closed-form expression for $\overline{\bm{{v_m^{c,\pm}}}}$, its algebraic complexity renders any analytical approach to study its properties impracticable.

The desert steady state $\overline{\bm{{v_m^d}}}$ is always linearly stable (the eigenvalues of its Jacobian are $-B_1$, $-B_2$, $-1$). The grass equilibrium $\overline{\bm{{v_m^{g,+}}}}$ is linearly stable for 

\begin{align*}
A<A_u^{G} := \frac{B_2^2 + k_1^2\left(B_2-FB_1\right)^2}{Fk_1\left(B_2-FB_1\right)},
\end{align*}	
provided $0<B_2-FB_1<FB_1$ and $k_1>\sqrt{B_2(2FB_1-B_2)}(B_2-FB_1)^{-1}$, and unstable otherwise. The second grass equilibrium $\overline{\bm{{v_m^{g,-}}}}$ is unstable. The tree equilibrium $\overline{\bm{{v_m^{t,+}}}}$ is stable for 

\begin{align*}
A<A_u^{T} := \frac{F^2B_1^2 + Hk_2^2\left(B_2-FB_1\right)^2}{FHk_2\left(FB_1-B_2\right)},
\end{align*}
provided $-B_2<B_2-FB_1<0$ and $k_2>\sqrt{B_1FH(2B_2-FB_1)}(H(FB_1-B_2))^{-1}$, and unstable otherwise. The second tree equilibrium $\overline{\bm{{v_m^{t,-}}}}$ is unstable. Existence and stability of the coexistence equilibria $\overline{\bm{{v_m^{c,\pm}}}}$ are found using the numerical continuation software AUTO-07p \cite{AUTO}. The lower branch $\overline{\bm{{v_m^{c,-}}}}$ is always unstable, while $\overline{\bm{{v_m^{c,+}}}}$ is stable if intraspecific competition is sufficiently stronger than interspecific competition. In particular, the {\color{changes}local} intraspecific competition of the locally superior species needs to be sufficiently strong for coexistence to be stable, while that of the locally inferior species only has a negligible effect on the stability of the equilibrium.

The upper bounds on the rainfall parameter and other constraints required for stability of the spatially uniform single-species equilibria are a crucial difference to the stability results for the single-species model \eqref{eq: Multispecies pattern: single-species model}. As precipitation is increased, the single-species equilibria lose their stability to the coexistence equilibrium $\overline{\bm{{v_m^{c,+}}}}$, because an increase in resource availability causes a reduction in the strength of interspecific competition (Fig. \ref{fig: Multispecies pattern: spatially uniform bifdiag}). In the absence of {\color{changes}local} intraspecific competition, no coexistence equilibrium exists and no upper bound on the rainfall parameter for stability of the single-species equilibria exists. 

Moreover, both in \eqref{eq: Multispecies pattern: Model: nondimensional model} and in the absence of {\color{changes}local} intraspecific competition, no bistability of the single-species equilibria can occur, as the upper precipitation bounds satisfy $A_u^g A_u^t <0$ (Fig. \ref{fig: Multispecies pattern: spatially uniform bifdiag}). The quantity $B_2-FB_1$, which determines the signs of $A_u^g$ and $A_u^t$, denotes the local average fitness difference between both species in the absence of any {\color{changes}local} intraspecific competition \cite{Eigentler2019Multispecies}. A definition of local average fitness in \eqref{eq: Multispecies pattern: Model: nondimensional model} is not as straightforward as in the model with no {\color{changes}local} intraspecific competition, but the stability thresholds $A_u^G$ and $A_u^T$ highlight that {\color{changes}local} intraspecific competition cannot change which species is of higher local average fitness.

\subsection{Single-species patterns}

Onset and existence of single-species patterns remain independent of the introduction of a second species, i.e. results presented for the single species model \eqref{eq: Multispecies pattern: single-species model} also hold for the multispecies model \eqref{eq: Multispecies pattern: Model: nondimensional model}. By contrast, stability of single-species patterns is significantly affected by the introduction of a competitor species and is also related to the onset of coexistence patterns.

As for the single species model \eqref{eq: Multispecies pattern: single-species model}, patterned solutions of \eqref{eq: Multispecies pattern: Model: nondimensional model} are limit cycles of the corresponding travelling wave ODE system

\begin{subequations}\label{eq: Multispecies pattern: single-species patterns: tw model}
	\begin{align}
	WU_1\left(U_1 + HU_2\right)\left(1-\frac{U_1}{k_1}\right) - B_1 U_1 + c\frac{\dif U_1}{\dif z} +\frac{\dif^2 U_1}{\dif z^2} &=0,\label{eq: Multispecies: Model: tw model u1} \\
	FWU_2\left(U_1 + HU_2\right)\left(1-\frac{U_2}{k_2}\right) - B_2 U_2 + c\frac{\dif U_2}{\dif z} +D\frac{\dif^2 U_2}{\dif z^2}&=0, \label{eq: Multispecies: Model: tw model u2}\\
	A-W - W\left(U_1+U_2\right)\left(U_1 + HU_2\right) + (c+\nu) \frac{\dif W}{\dif z} +d \frac{\dif^2 W}{\dif z^2} &=0,
	\end{align}
\end{subequations}
which is obtained from the PDE model \eqref{eq: Multispecies pattern: Model: nondimensional model} by setting  $u_1(x,t) = U_1(z)$, $u_2(x,t) = U_2(z)$ and $w(x,t) = W(z)$ for $z=x-ct$, $c \in \R$. As in the single-species model \eqref{eq: Multispecies pattern: single-species model}, this introduces a new parameter, the uphill migration speed $c$, and the bifurcation analysis is performed in the $(A,c)$ parameter plane. However, for illustrative purposes, I fix the migration speed in the presentation of the bifurcation diagrams, but emphasise that the results do not qualitatively depend on the choice of $c$, unless otherwise stated. The transformation into the travelling wave framework enables the calculation of a pattern's essential spectrum to determine its stability using the numerical continuation method by Rademacher et al. \cite{Rademacher2007}, and I again refer to \cite{Rademacher2007,Sherratt2012,Sherratt2013a} for full details on the method and to \cite{Eigentler2020savanna_coexistence} for an overview on how this algorithm is implemented for \eqref{eq: Multispecies pattern: Model: nondimensional model} in the limit $k_1,k_2 \rightarrow \infty$.

\begin{figure}
	\centering
	\subfloat[\label{fig: Multispecies pattern: spectra comparison single}]{\includegraphics[width=0.48\textwidth]{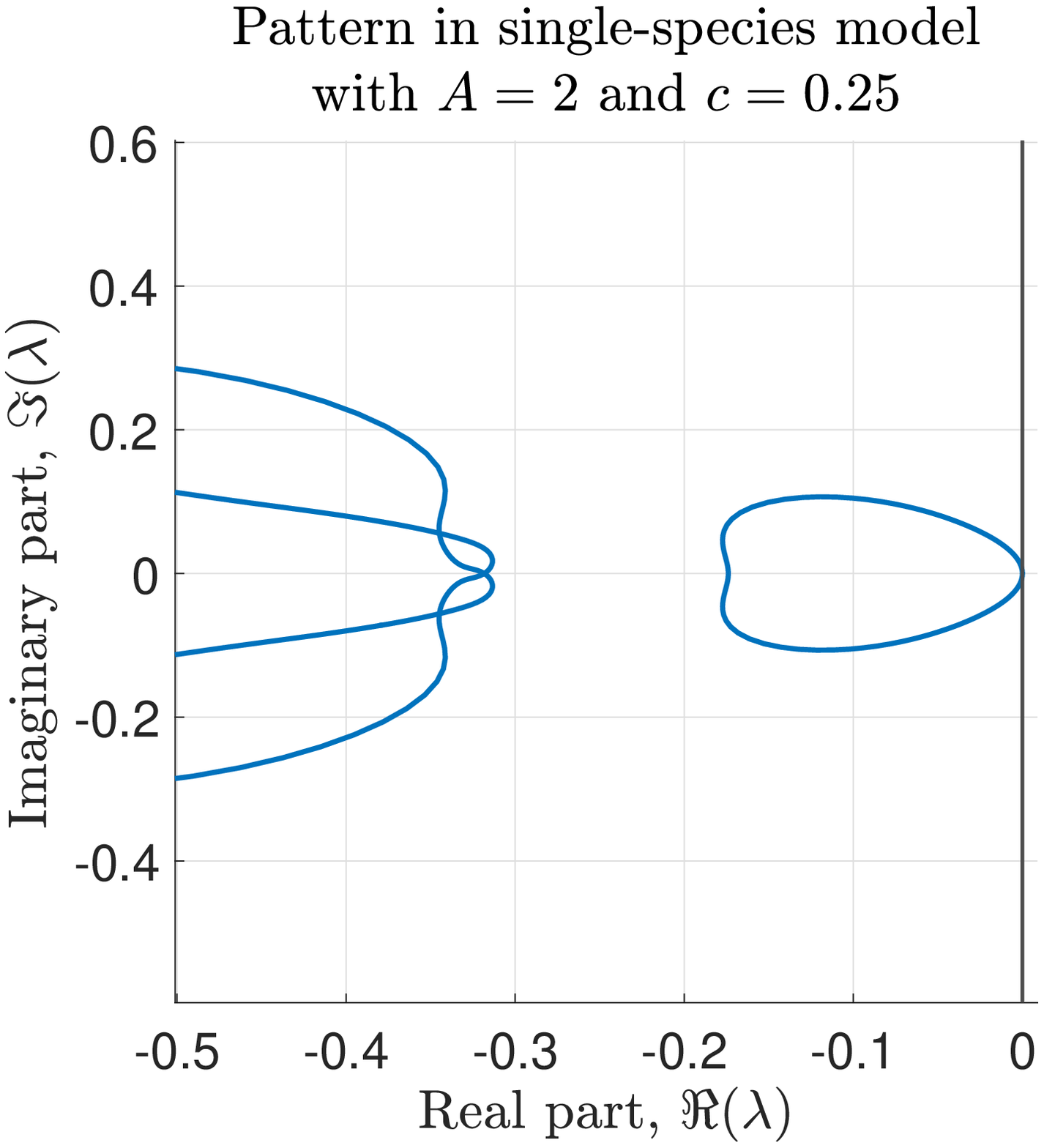}}
	\subfloat[\label{fig: Multispecies pattern: spectra comparison multi}]{\includegraphics[width=0.48\textwidth]{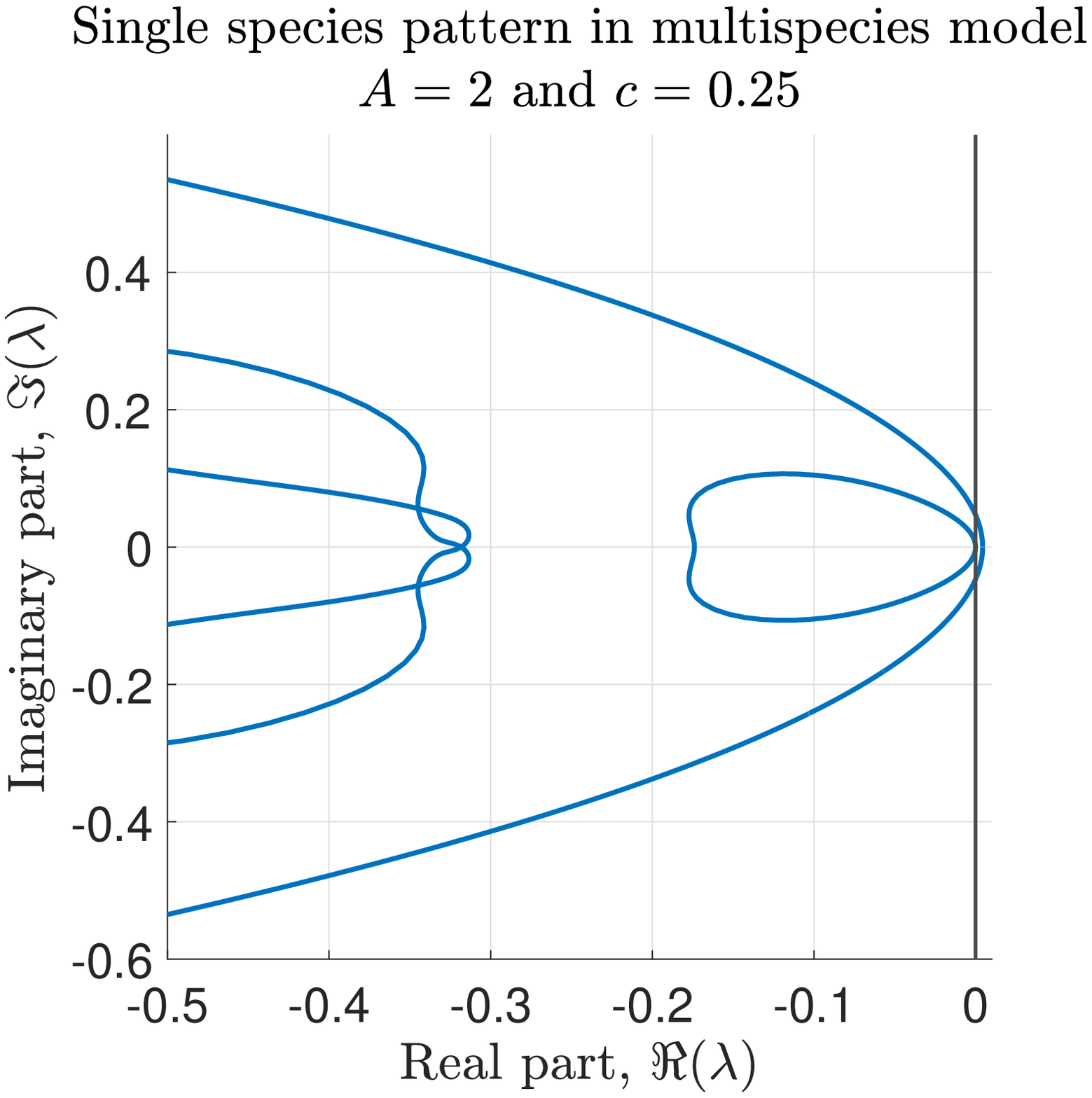}}
	\caption{\textbf{Introduction of a second species affects stability of single-species patterns.} A comparison of the essential spectra of a single-species pattern in the single-species model \eqref{eq: Multispecies pattern: single-species model} (a) and the multispecies model \eqref{eq: Multispecies pattern: Model: nondimensional model} (b) are shown. The spectrum in the single-species model is a subset of the spectrum in the multispecies model. The additional elements of the spectrum correspond to the leading order behaviour of perturbations {\color{changes}in the density of the second species.} Note that the spectra yield that the corresponding single-species pattern is stable in the single-species model, but unstable in the multispecies model due to the introduction of {\color{changes} the competitor species}. The vertical lines visualise the imaginary axis. The parameter values are $A=2$ and $c=0.25$. {\color{changes}For this visualisation, a pattern of species $u_1$ was chosen, but identical considerations hold for single-species patterns of species $u_2$.}} \label{fig: Multispecies pattern: spectra comparison}
\end{figure}

Unlike pattern onset and existence, the stability of single-species patterns of \eqref{eq: Multispecies pattern: Model: nondimensional model} is affected by the second species in the system. For a single-species pattern to be stable in the multispecies model \eqref{eq: Multispecies pattern: Model: nondimensional model}, it needs to be stable in the context of the single-species model \eqref{eq: Multispecies pattern: single-species model} and stable to the introduction of the competitor species, two conditions that are independent of each other. The stability of a single-species pattern to the introduction of the competitor species is determined by a comparison of its essential spectrum in the multispecies model with that of the same solution in the single-species model (Fig. \ref{fig: Multispecies pattern: spectra comparison}). The spectrum of the periodic travelling wave in the single-species model is a subset of that of the solution in the multispecies model. The additional elements in the latter describe the leading order behaviour of perturbations due to the introduction of the competitor species. Thus, a pattern that is stable in the corresponding single-species model may be unstable in the multispecies model \eqref{eq: Multispecies pattern: Model: nondimensional model} due to its interaction with a competitor species. 

\subsection{Onset and existence of coexistence patterns}\label{sec: Multispecies pattern: onset and existence}

Onset of coexistence patterns can occur through three different mechanisms. As for the single-species patterns discussed in Sec. \ref{sec: Multispecies pattern: single-species}, two potential causes of pattern onset are a homoclinic solution and a Turing-Hopf bifurcation of the spatially uniform coexistence equilibrium $\bm{{v_m^{c,+}}}$. Onset of coexistence patterns can further occur on a solution branch of a single-species pattern as it loses/gains stability to the introduction of the second species. As outlined in the previous section, such a bifurcation can be detected through a comparison of the single-species pattern's essential spectra in the context of the single-species model \eqref{eq: Multispecies pattern: single-species model} and the multispecies model \eqref{eq: Multispecies pattern: Model: nondimensional model}. The same mechanism also causes pattern onset {\color{changes}if only intraspecific competition for water is considered} \cite{Eigentler2020savanna_coexistence}. Onset at a homoclinic solution or at a Turing-Hopf bifurcation of a spatially uniform equilibrium, however, cannot occur if {\color{changes}local} intraspecific competition dynamics are neglected, as no spatially uniform equilibria exist. In \eqref{eq: Multispecies pattern: Model: nondimensional model}, solution branches of coexistence patterns either connect two single-species patterns (the only mechanism that occurs in the absence of {\color{changes}local} intraspecific competition), a single-species pattern with the spatially uniform coexistence state, or the spatially uniform coexistence state with a homoclinic solution. The choice of which of these three mechanisms occurs depends on both the strength of {\color{changes}local} intraspecific competition and the difference between both species, as is outlined below.

\subsubsection{The role of {\color{changes}local} intraspecific competition}

\begin{figure}
	\centering
	\subfloat[$k_1=k_2=10$ (strong {\color{changes}local} intraspecific competition among both species)\label{fig: Multispecies pattern: bifurcation diag k small}]{\includegraphics[width=0.3\textwidth, trim = 0cm 0cm 0cm 1.2cm, clip]{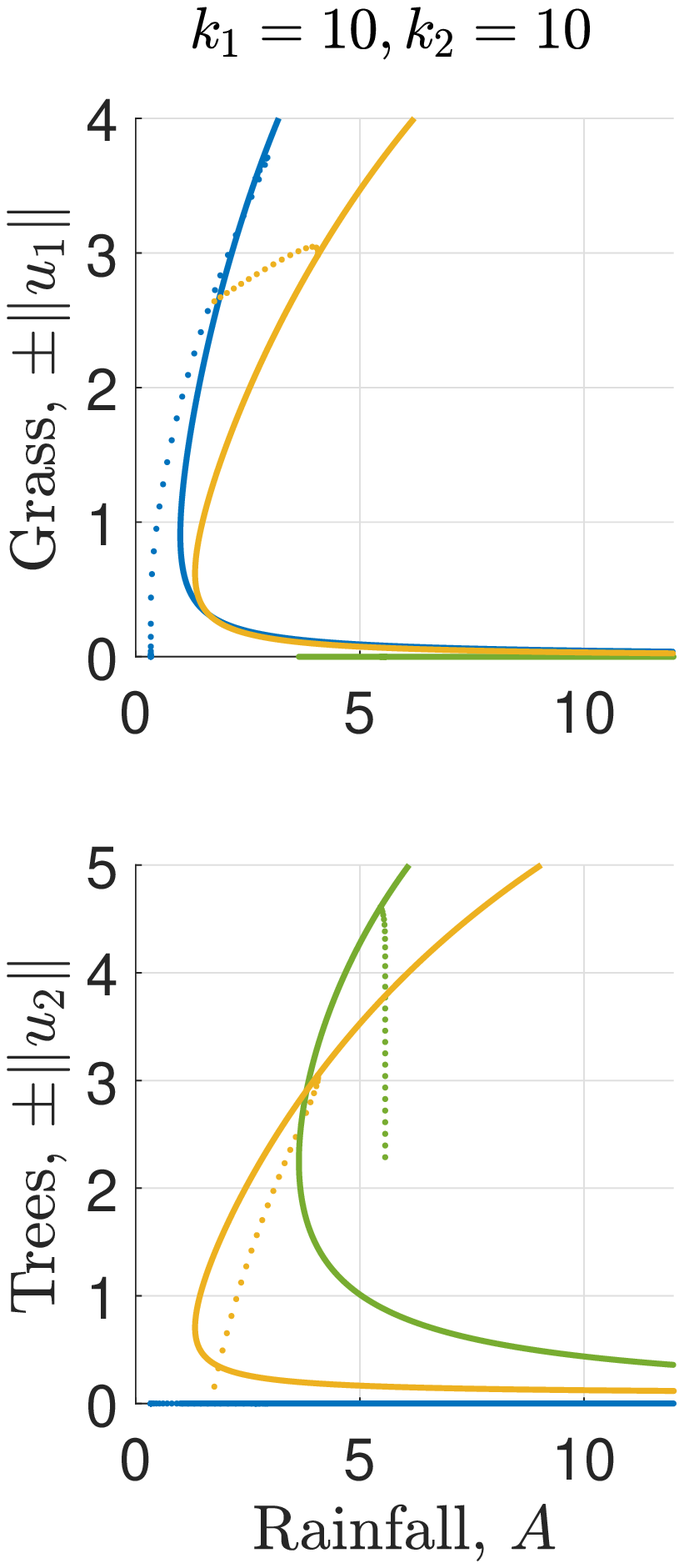}} \hfill
	\subfloat[$k_1=k_2=2000$ (intermediate {\color{changes}local} intraspecific competition among both species)\label{fig: Multispecies pattern: bifurcation diag k medium}]{\includegraphics[width=0.3\textwidth, trim = 0cm 0cm 0cm 1.2cm, clip]{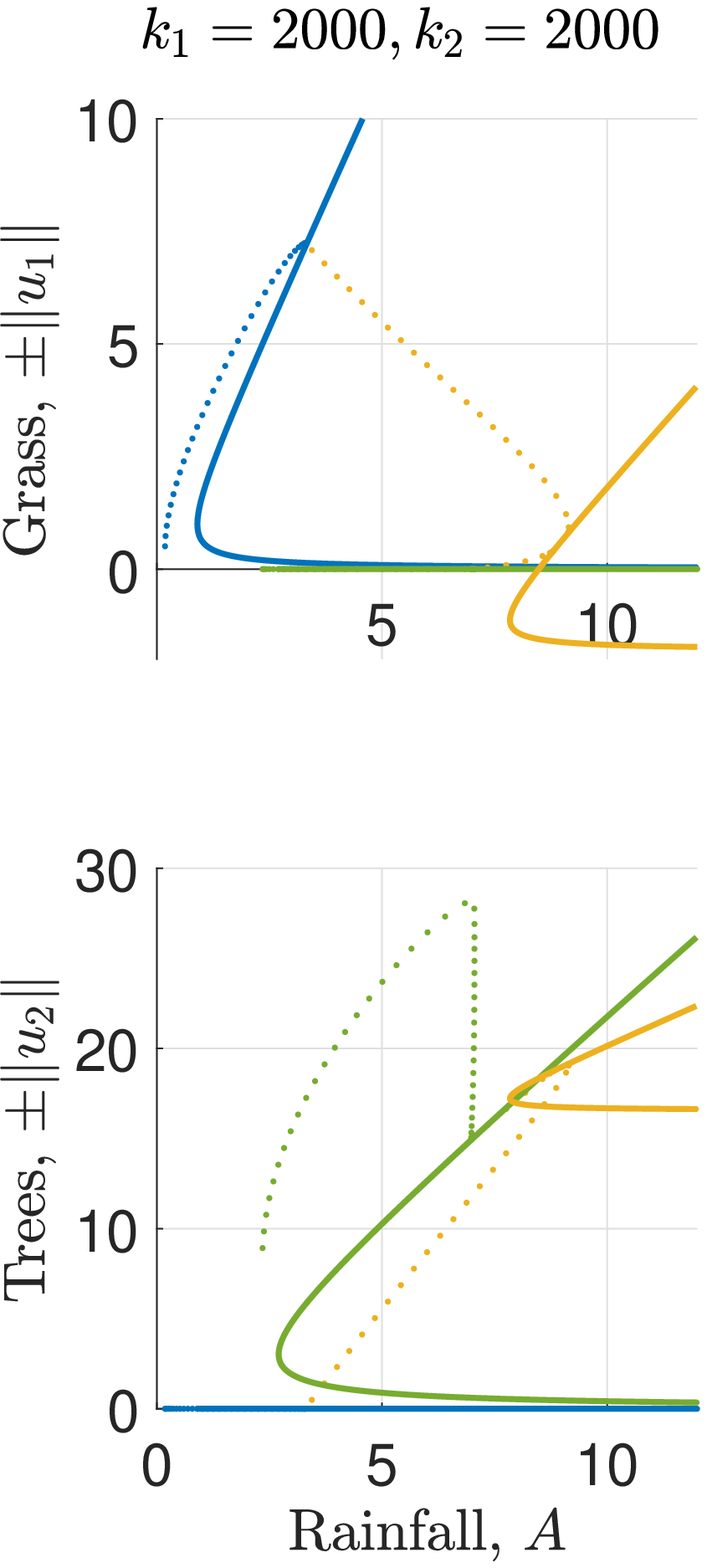}} \hfill
	\subfloat[$k_1=k_2=2100$ (weak {\color{changes}local} intraspecific competition among both species)\label{fig: Multispecies pattern: bifurcation diag k large}]{\includegraphics[width=0.3\textwidth, trim = 0cm 0cm 0cm 1.2cm, clip]{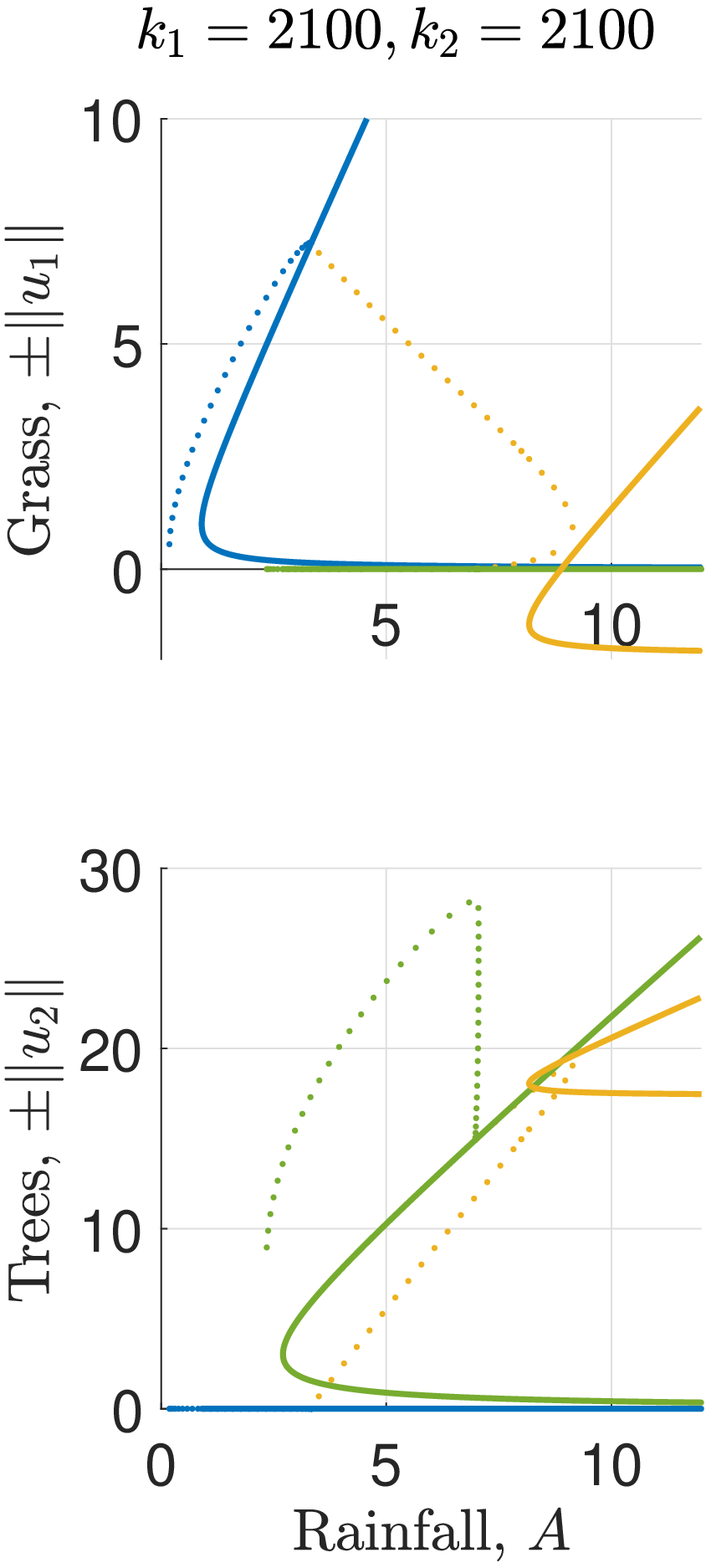}} \\
	\begin{minipage}{0.3\textwidth}
		\includegraphics[width=\textwidth]{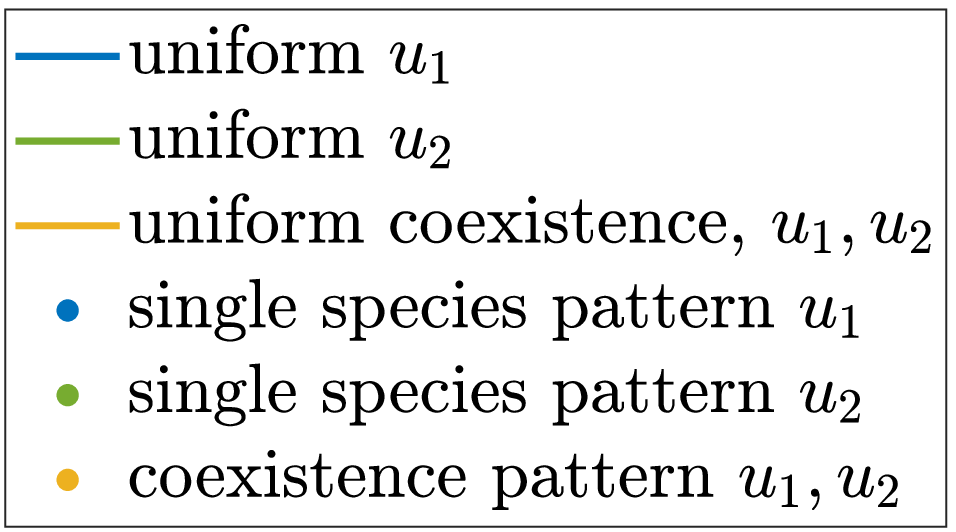}
	\end{minipage}
	\hfill
	\begin{minipage}{0.68\textwidth}
		\caption{\textbf{Strong {\color{changes}local} intraspecific competition facilitates spatially uniform coexistence and causes coexistence pattern onset at a Turing-Hopf bifurcation.} Bifurcation diagrams for different values of the carrying capacities $k_1=k_2$ are shown for $c=0.25$. A decrease in {\color{changes}local} intraspecific competition increases the size of the precipitation interval in which coexistence patterns exist and simultaneously inhibits spatially uniform coexistence. Under strong {\color{changes}local} intraspecific competition, two Hopf bifurcations along the spatially uniform coexistence equilibrium exist and cause the onset of patterns. Typically, patterns originating from the lower branch are of large wavelength and are thus omitted form the bifurcation diagram in (a). Both Hopf bifurcation loci meet in a fold as {\color{changes}local} intraspecific competition is increased to a critical threshold beyond which coexistence patterns connect both single-species pattern branches ((b) and (c)). Patterned states are only shown for one value of the uphill migration speed and no stability information is provided. In (b) and (c), $\|u_1\|$ is multiplied by $\operatorname{sign}(u_1)$ to visualise the occurrence of $u_1<0$. }\label{fig: Multispecies pattern: bifurcation diag k} 
	\end{minipage}
\end{figure}

If $k_1=k_2$ is small and species difference is sufficiently large so that $u_1$ and $u_2$ represent a typical grass and tree species, respectively, two Hopf bifurcations on the spatially uniform coexistence equilibria occur and are the origins of coexistence pattern solution branches that connect to either of the single-species pattern branches. (Fig. \ref{fig: Multispecies pattern: bifurcation diag k small}). Typically, one of the Hopf bifurcations occurs on $\bm{{v_m^{c,-}}}$ and patterns originating there are of very large wavelength, beyond the $L=1000$ threshold used to approximate homoclinic solutions in this bifurcation analysis. Note that the Hopf bifurcation on $\bm{{v_m^{c,-}}}$ does not cause a stability change of the equilibrium because a third eigenvalue with positive real part exists. As $k_1=k_2$ increases, the spatially uniform coexistence equilibrium is shifted to higher precipitation volumes and one of its biomass components may attain ecologically irrelevant negative values. Moreover, the Hopf bifurcation on $\bm{{v_m^{c,-}}}$ moves along the solution branch, through the fold, and onto the $\bm{{v_m^{c,+}}}$ branch (Fig. \ref{fig: Multispecies pattern: bifurcation diag k medium}). A further increase in $k_1=k_2$ reduces the distance between both Hopf bifurcations, until they coincide. Beyond this threshold, no Hopf bifurcation along the spatially uniform coexistence equilibrium exists. However, coexistence patterns continue to occur. As in the analysis shown in \cite{Eigentler2020savanna_coexistence} (the $k_1, k_2\rightarrow\infty$ limit of the model in this paper), one coexistence pattern solution branch connects both single-species pattern branches for sufficiently large $k_1=k_2$ (Fig. \ref{fig: Multispecies pattern: bifurcation diag k large}). In other words, {\color{changes}local} intraspecific competition shifts the existence region of both the spatially uniform coexistence equilibrium and the spatially patterned coexistence state to lower precipitation levels and enables coexistence pattern onset at a Hopf bifurcation on the spatially uniform equilibrium. 

\begin{figure}
	\centering
	\subfloat[$A=2$, $k_1=k_2=10$\label{fig: Multispecies pattern: increase k and A low A k plot}]{\includegraphics[width=0.48\textwidth]{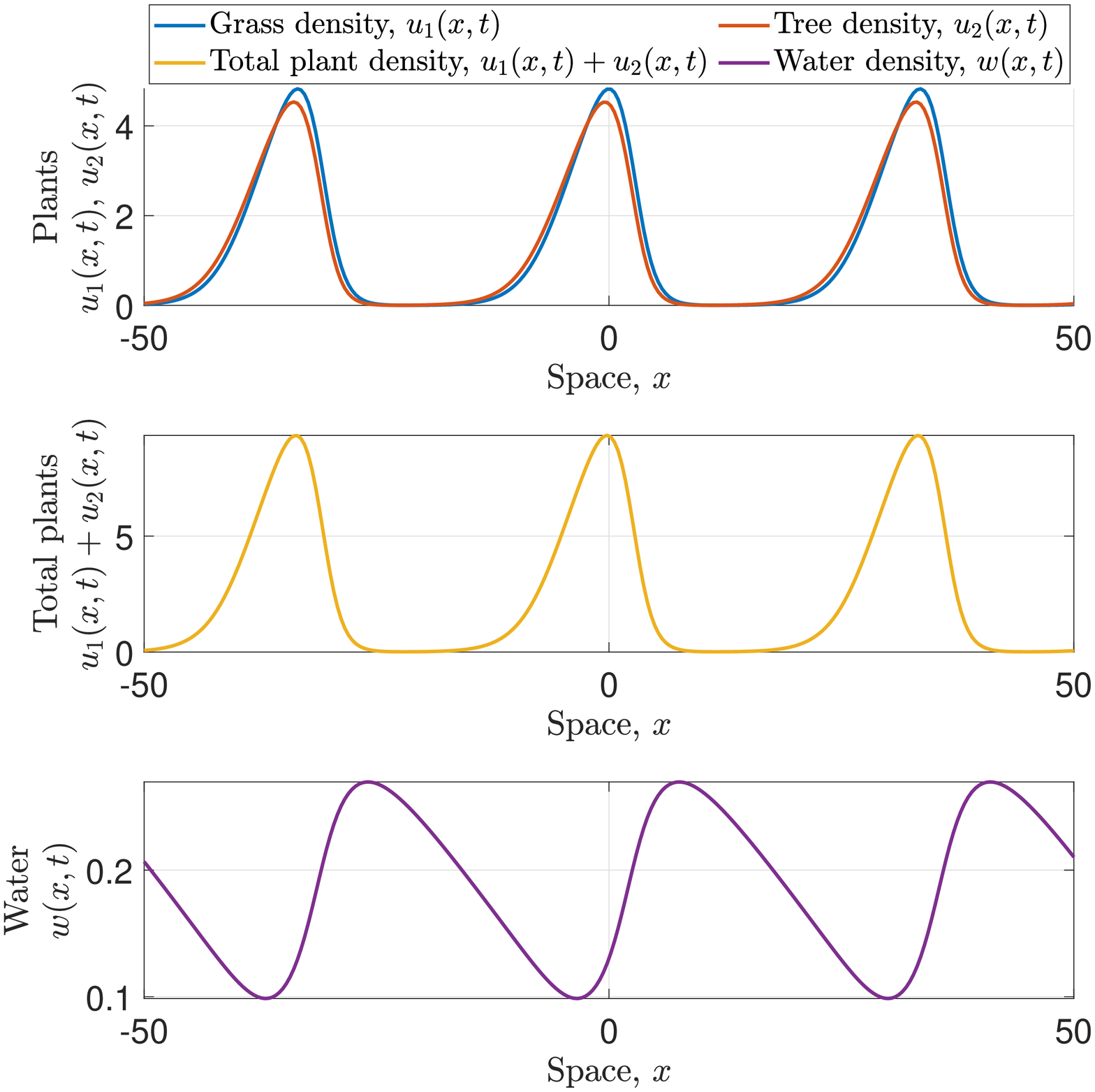}}	\subfloat[$A=3.75$, $k_1=k_2=100$\label{fig: Multispecies pattern: increase k and A high A k plot}]{\includegraphics[width=0.48\textwidth]{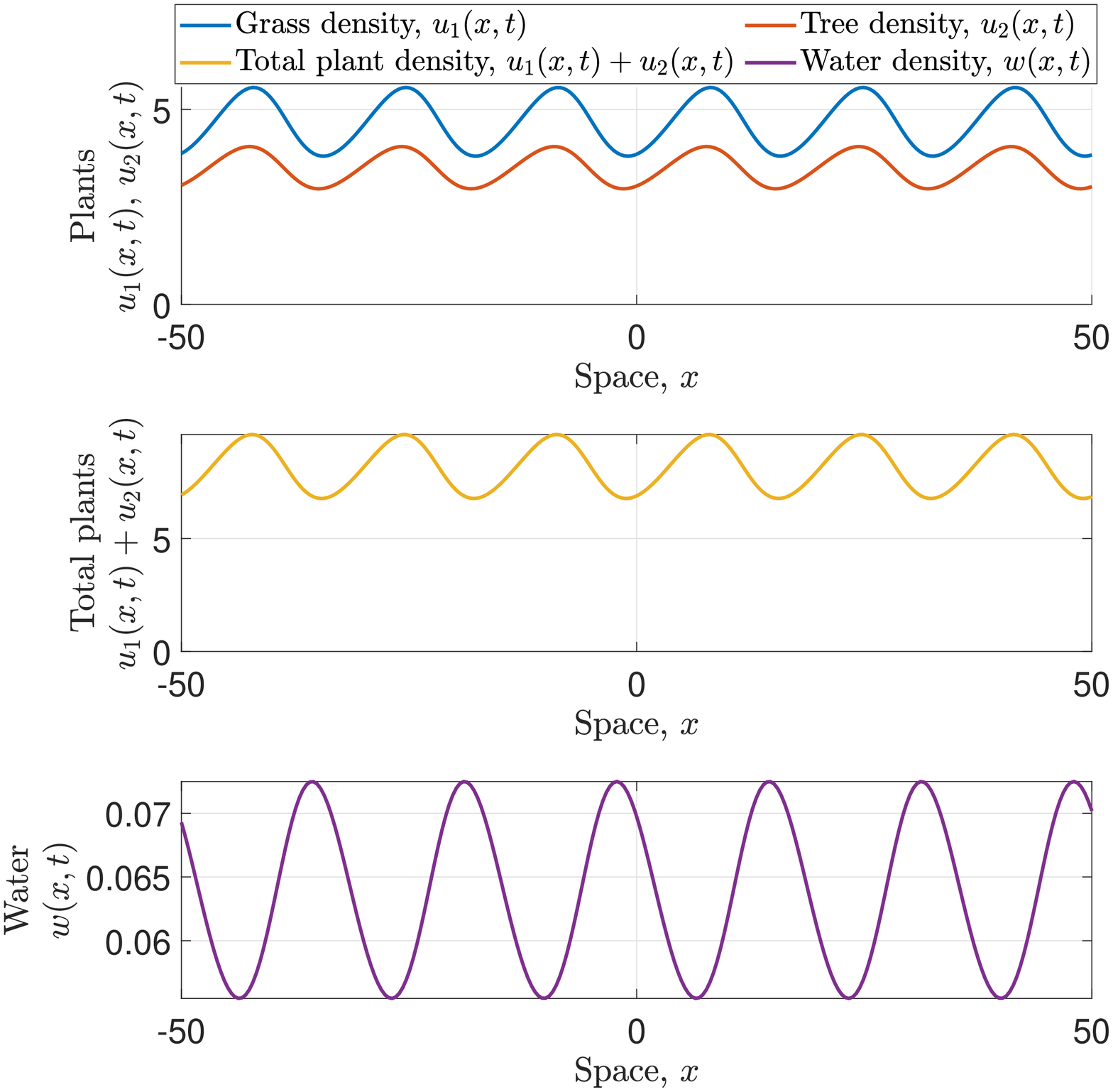}}\\
	\caption{\textbf{{\color{changes}Local} intraspecific competition facilitates species coexistence in vegetation patterns.} Two coexistence solutions are shown. In (a), {\color{changes}local} intraspecific competition is strong and the solution represents a vegetation pattern, while in (b) a solution corresponding to a savanna state is visualised, which occurs due to weak {\color{changes}local} intraspecific competition. Note the different values of the precipitation parameter. A decrease in {\color{changes}local} intraspecific competition destabilises the coexistence state at lower rainfall volumes. The species difference parameter is $\chi=0.3$.}\label{fig: Multispecies pattern: increase k and A}	
\end{figure}

\begin{figure}
	\centering
	\subfloat[$k_1 = 10, k_2 = 1500$ (strong {\color{changes}local} intraspecific competition among the coloniser species, intermediate {\color{changes}local} intraspecific competition among the locally superior species)\label{fig: Multispecies pattern: bifurcation diag k2 medium} ]{\includegraphics[width=0.3\textwidth, trim = 0cm 0cm 0cm 1.2cm, clip]{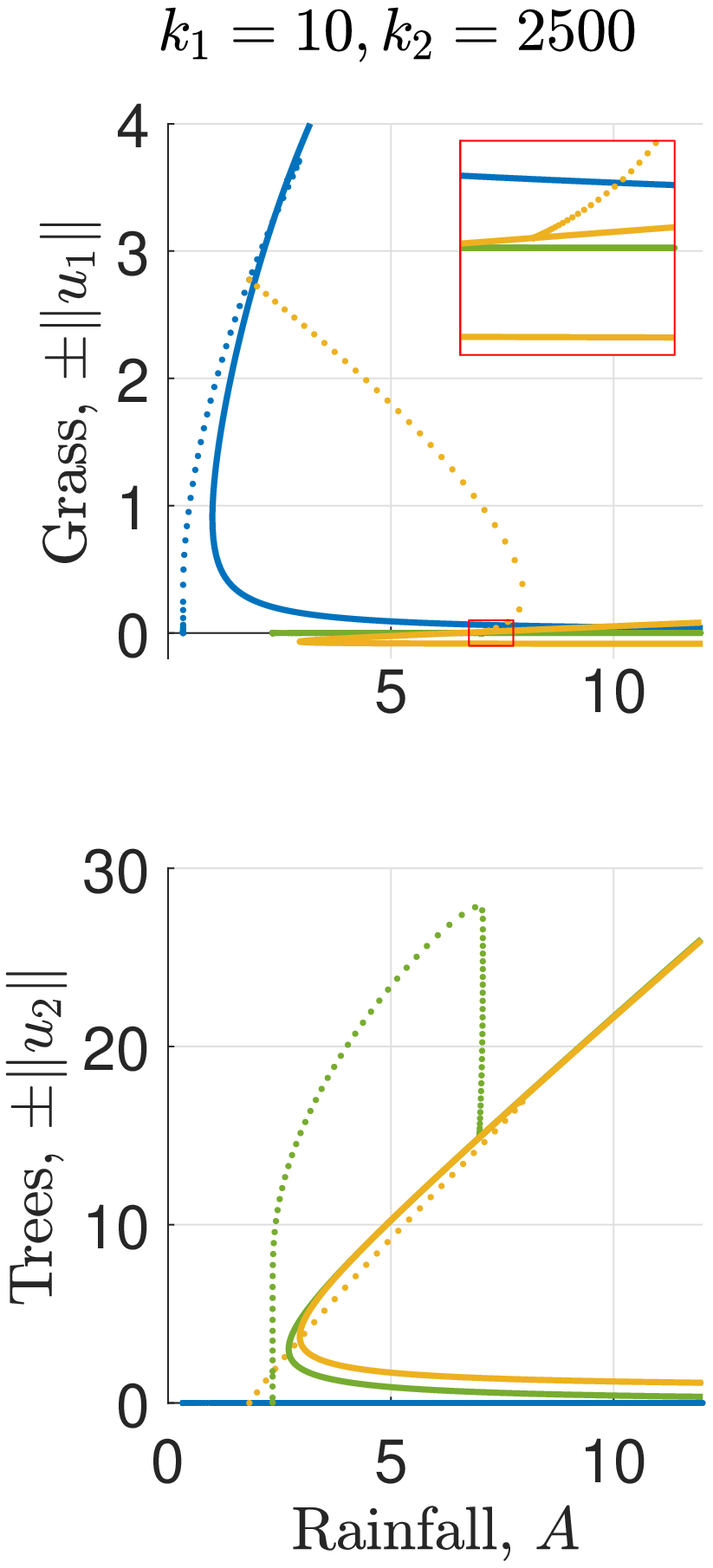}} \hfill
	\subfloat[$k_1 = 10, k_2 = 2500$ (strong {\color{changes}local} intraspecific competition among the coloniser species, weak {\color{changes}local} intraspecific competition among the locally superior species)\label{fig: Multispecies pattern: bifurcation diag k2 large} ]{\includegraphics[width=0.3\textwidth, trim = 0cm 0cm 0cm 1.2cm, clip]{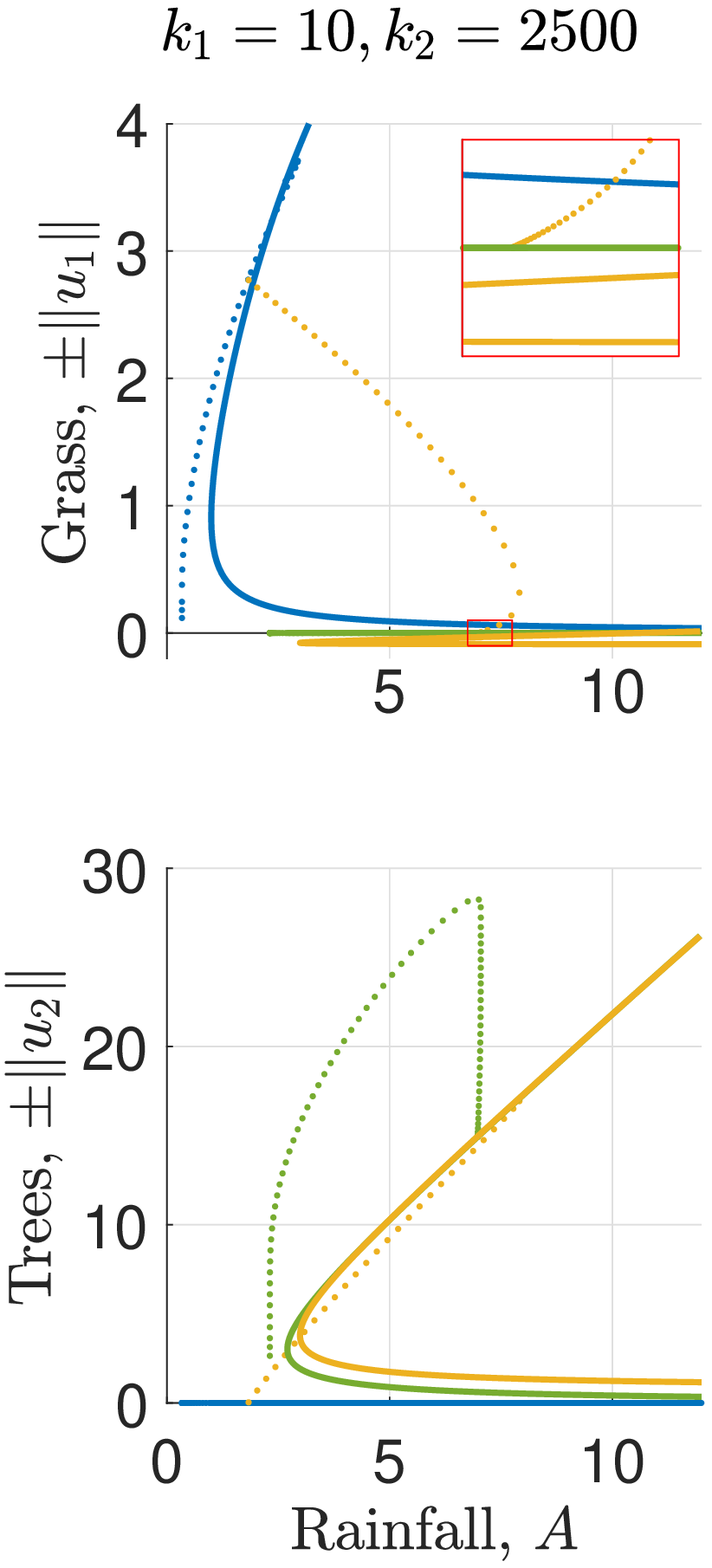}} \hfill
	\subfloat[$k_1 = 1000, k_2 = 10$ (weak {\color{changes}local} intraspecific competition among the coloniser species, strong {\color{changes}local} intraspecific competition among the locally superior species)\label{fig: Multispecies pattern: bifurcation diag k1 large} ]{\includegraphics[width=0.3\textwidth, trim = 0cm 0cm 0cm 1.2cm, clip]{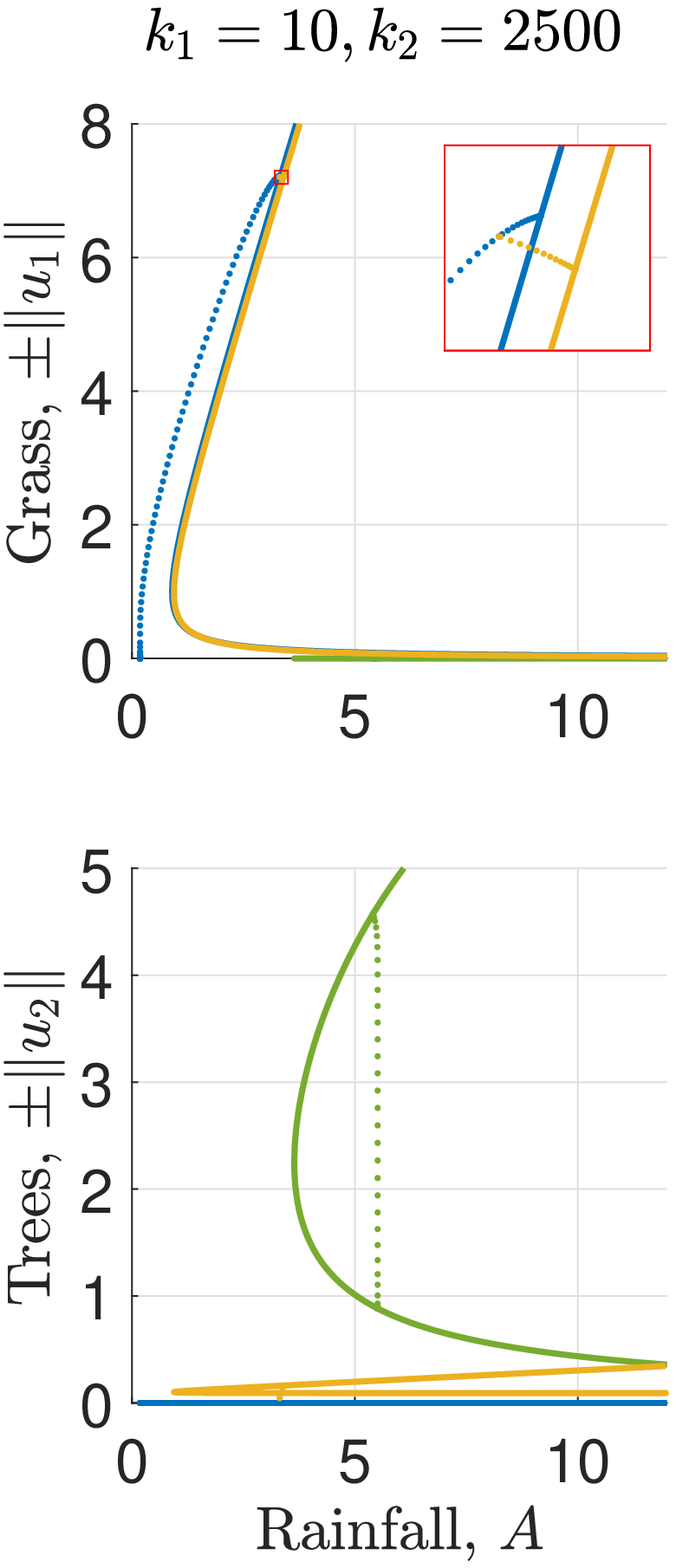}}
	\caption{\textbf{Strong {\color{changes}local} intraspecific competition of the coloniser species and weak {\color{changes}local} intraspecific competition of the locally superior species promote patterned coexistence.} Bifurcation diagrams under changing {\color{changes}local} intraspecific competition of one-species only are shown. Both strong {\color{changes}local} intraspecific competition among the coloniser species $u_1$ and weak {\color{changes}local} intraspecific competition among the locally superior species $u_2$ increase the size of the parameter region in which coexistence patterns exist. The insets in (a) and (b) (axes limits: $A\in [6.75,7.75]$, $\pm\|u_1\| \in [-0.1,0.1]$)  show the onset of coexistence patterns close to $u_1=0$ to highlight the transition from onset at the spatially uniform coexistence equilibrium to onset at the single-species $u_2$ pattern as {\color{changes}local} intraspecific competition among $u_2$ decreases. The inset in (c) (axes limits: $A\in [3.2,3.5]$, $\pm\|u_1\| \in [7.1,7.3]$)  shows a blow-up of the parameter region in which coexistence pattern exist. The pattern migration speed is $c=0.25$. In (a) and (b), $\|u_1\|$ is multiplied by $\operatorname{sign}(u_1)$ to visualise the occurrence of $u_1<0$. For an interpretation of colours and linestyles used in the visualisation, see the legend of Fig. \ref{fig: Multispecies pattern: bifurcation diag k}. }\label{fig: Multispecies pattern: bifurcation diag ks separate} 
\end{figure}

An investigation with one of the species' {\color{changes}local} intraspecific competition strengths being fixed, gives more insight into the different roles of both parameters. A decrease in {\color{changes}local}  intraspecific competition of the coloniser species (i.e. increase in $k_1$) reduces the size of the parameter region for which coexistence patterns occur (Fig. \ref{fig: Multispecies pattern: bifurcation diag k1 large}). As is discussed in \cite{Eigentler2020coexistence_pattern} and visualised in Fig. \ref{fig: Multispecies pattern: bifurcation diag k2 medium} and \ref{fig: Multispecies pattern: bifurcation diag k2 large}, strong {\color{changes}local} intraspecific competition among the coloniser species facilitates coexistence patterns because it shifts the upper rainfall threshold at which pattern onset occurs to higher levels, while only having a negligible impact on the onset at low precipitation volumes. This causes an increase in the size of the parameter region in which coexistence patterns exist. Variations in $k_2$, however, have a very similar effect as in the case of $k_1=k_2$ (Fig. \ref{fig: Multispecies pattern: bifurcation diag k2 medium} and \ref{fig: Multispecies pattern: bifurcation diag k2 large}). A reduction in {\color{changes}local} intraspecific competition increases the size of the pattern existence region. In contrast to the $k_1=k_2$ case, the Hopf bifurcation on the lower branch of the spatially uniform coexistence equilibrium has no impact on the structure of ecologically relevant solutions, as it exclusively occurs for parameter values at which one of the plant densities of the coexistence equilibrium is negative. Nevertheless, a transition to a bifurcation structure in which the coexistence pattern solution branch connects both single-species patterns occurs as follows. As $k_2$ increases the $u_1$ density of the spatially uniform coexistence equilibrium decreases and becomes negative after intersecting the single-species tree equilibrium. Consequently, the Hopf bifurcation on the equilibrium occurs for lower densities of $u_1$ as $k_2$ increases (Fig. \ref{fig: Multispecies pattern: bifurcation diag k2 medium}). At a critical threshold, the Hopf bifurcation crosses $u_1=0$, where it coincides with the Hopf bifurcation on the single-species tree equilibrium. For $k_2$ larger than this threshold, ecologically relevant patterns connect the Hopf bifurcations on the single-species equilibria and do not extend to the Hopf bifurcation on the coexistence equilibrium solution branch, as this occurs for $u_1<0$ (Fig. \ref{fig: Multispecies pattern: bifurcation diag k2 large}).

\subsubsection{Transition from a savanna to a patterned vegetation state}
Strong {\color{changes}local} intraspecific competition also changes the solution behaviour by facilitating species coexistence in a state representing vegetation patterns. As discussed above, increases in {\color{changes}local} intraspecific competition strength shift the parameter interval in which coexistence patterns occur to lower precipitation volumes (Fig. \ref{fig: Multispecies pattern: bifurcation diag k}). Associated with this is a transition from a solution-type that represents a savanna biome to a solution type that represents a vegetation pattern. Both these solution types are periodic travelling waves, but the biomass components of the former oscillate between two non-zero levels, while those of the latter oscillate between a nonzero plant density and zero (Fig. \ref{fig: Multispecies pattern: increase k and A low A k plot} and \ref{fig: Multispecies pattern: increase k and A high A k plot}). In general, the transition between the two solution types is a gradual process. However, it may be accelerated by a destabilisation and associated change in wavelength of a pattern. The savanna state patterned solution also occurs in the $k_1,k_2 \rightarrow \infty$ limit as discussed in \cite{Eigentler2020savanna_coexistence}.

\subsubsection{The role of species difference}
\begin{figure}
	\centering
	\subfloat[$\chi = 0.3$ (small species difference)\label{fig: Multispecies pattern: bifurcation diag chi 03} ]{\includegraphics[width=0.3\textwidth, trim = 0cm 0cm 0cm 1.2cm, clip]{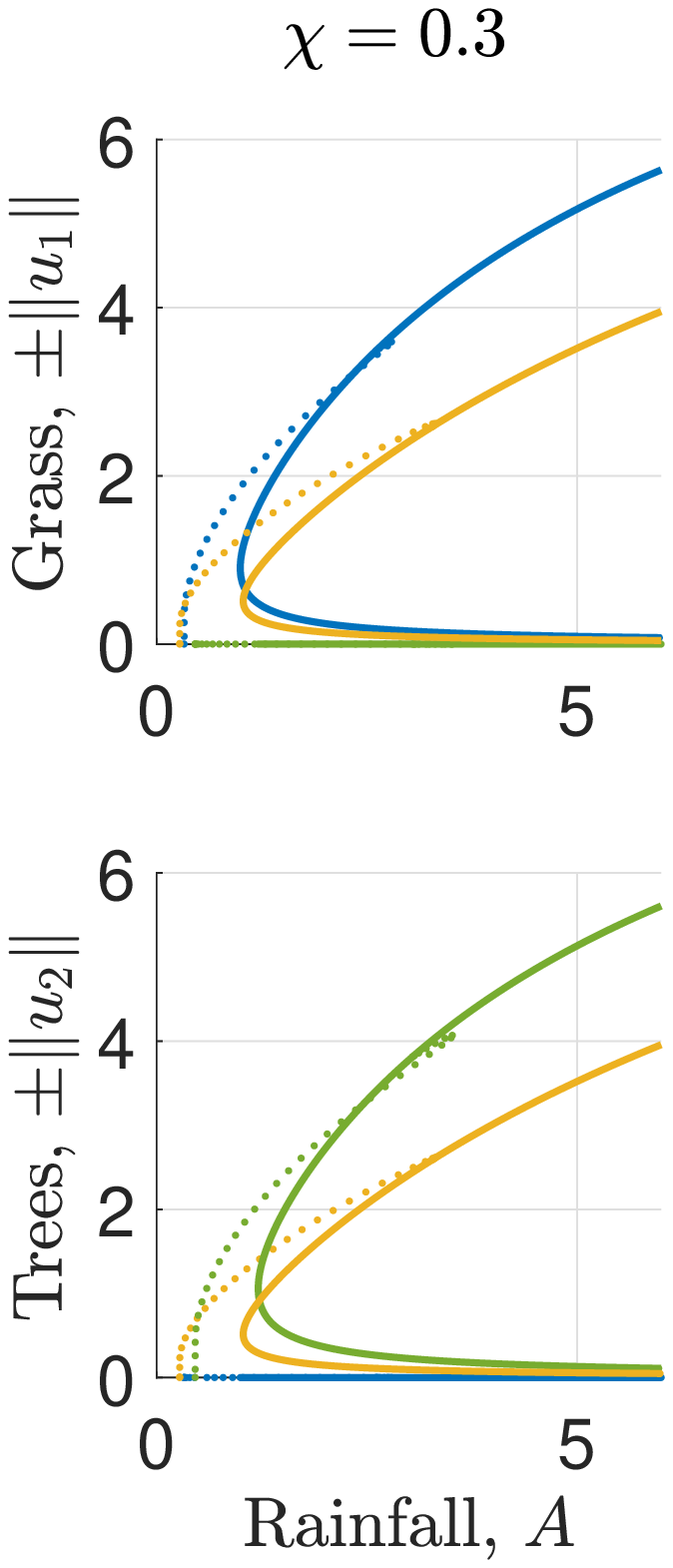}} \hfill
	\subfloat[$\chi = 0.9$ (large species difference)\label{fig: Multispecies pattern: bifurcation diag chi 09} ]{\includegraphics[width=0.3\textwidth, trim = 0cm 0cm 0cm 1.2cm, clip]{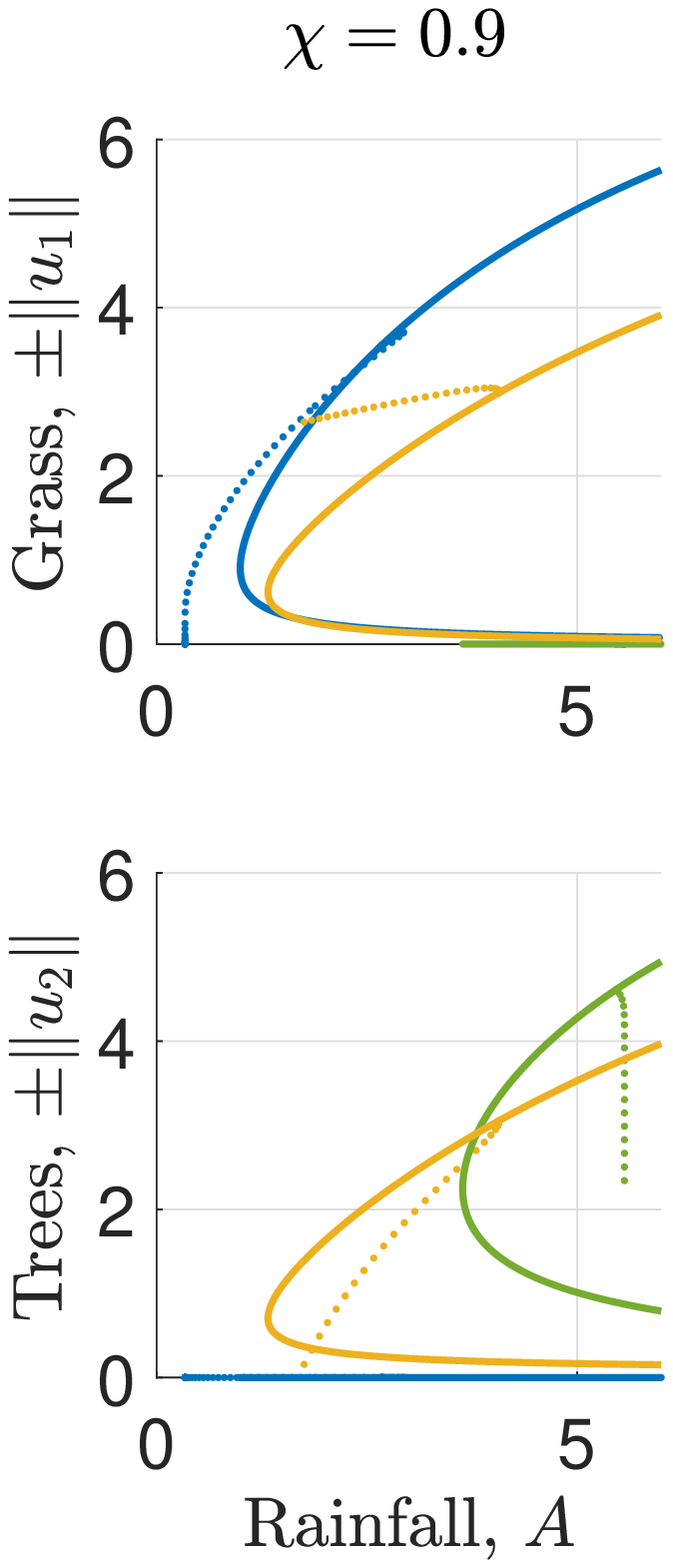}} \hfill
	\parbox{0.3\textwidth}{\vspace*{-5.1cm}\hfill\includegraphics[width=0.25\textwidth]{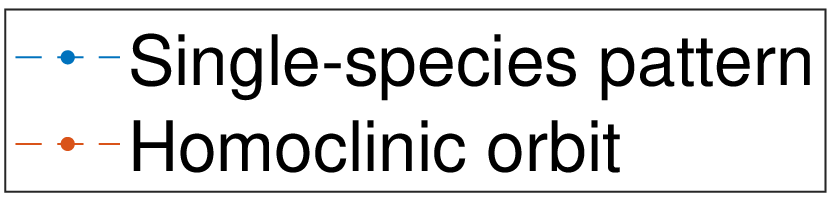} \\
		\subfloat[Pattern onset in ($A$,$\chi$) plane\label{fig: Multispecies pattern: bifurcation chi onset} ]{\includegraphics[width=0.3\textwidth]{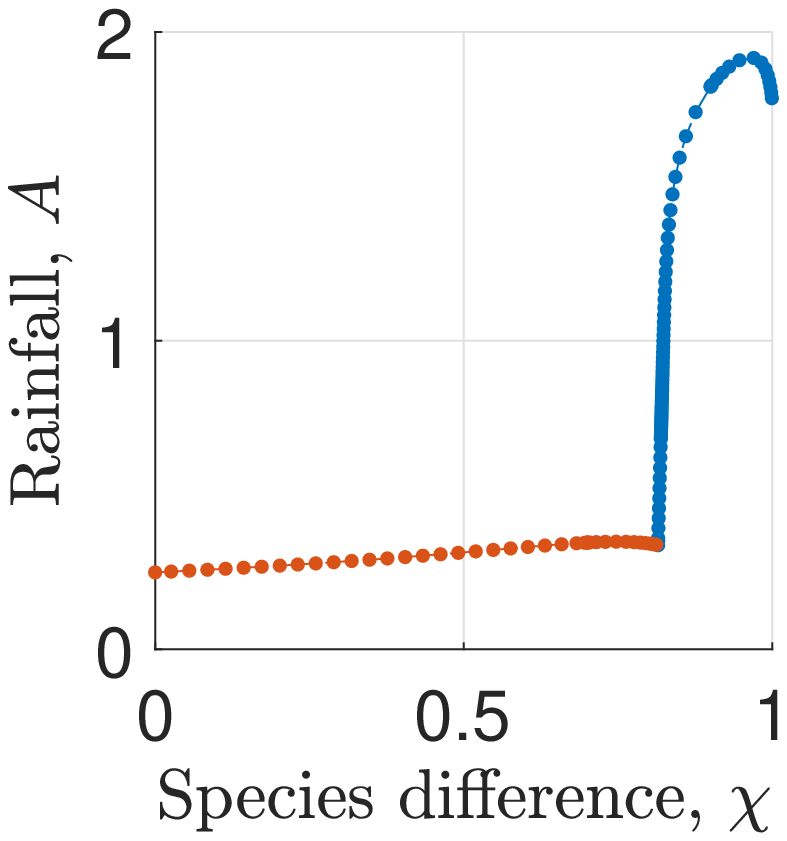}}
	}
	\caption{\textbf{A transition from coexistence pattern onset at a single-species pattern to onset at a homoclinic solution occurs due to increases in species similarity.} Bifurcation diagrams for different values of the species difference parameter $\chi$ are shown in (a) and (b). A transition from coexistence pattern onset at a homoclinic solution to onset at the single-species grass pattern occurs as species difference increases. The type of onset point and the precipitation level at which onset occur are tracked in (c). The pattern migration speed is $c=0.25$. For an interpretation of colours and linestyles used in (a) and (b), see the legend of Fig. \ref{fig: Multispecies pattern: bifurcation diag k}.}\label{fig: Multispecies pattern: bifurcation diag chi} 
\end{figure}

The difference between both plant species, quantified by the parameter $\chi$ in the parameter setting \eqref{eq: Multispecies pattern: parameter region to compare species}, also has a significant impact on the bifurcation structure of the system. In the results presented above, the difference between both species is set to a large value so that $u_1$ and $u_2$ represent a grass and tree species, respectively. Under this assumption, the onset of coexistence patterns at the lower precipitation bound for pattern existence always occurs along the single-species grass pattern. Decreases in the species difference $\chi$, corresponding to simultaneous changes in parameters of species $u_2$ that make it more similar to species $u_1$, cause the pattern onset locus to move along the single-species pattern branch in a decreasing precipitation direction towards the homoclinic solution of $u_1$. At a critical threshold of $\chi$, the homoclinic $u_1$ solution coincides with the homoclinic coexistence solution and a transition of the pattern onset type occurs. For lower values of the species difference parameter $\chi$, onset at low precipitation values thus occurs at the homoclinic solution (Fig. \ref{fig: Multispecies pattern: bifurcation diag chi}).

\subsection{The effects of plant dispersal}\label{sec: Multispecies pattern: phase diff}

As is discussed in \cite{Eigentler2020coexistence_pattern}, the ratio of the plant species' diffusion coefficients $D$ has a significant impact on the model solutions. Plant components of the patterned model solutions are not exactly in phase. Depending on the parameters in the system, the uphill edges (and to a lesser extent the downhill edges) of the travelling wave solutions are dominated by one species, while its competitor is mostly confined to narrow regions in the centre of the bands. This behaviour can be quantified through the linear correlation

\begin{align*}
\rho(U_1,U_2) = \frac{\operatorname{cov}(\widetilde{U_1},\widetilde{U_2})}{\sigma(\widetilde{U_1})\sigma(\widetilde{U_2})},
\end{align*}
between both plant densities, where $\operatorname{cov}(\cdot, \cdot)$ denotes the covariance of two vectors, and $\sigma(\cdot)$ the standard deviation. The vectors $\widetilde{U_1}$ and $\widetilde{U_2}$ are obtained by discretising the spatial domain and evaluating the plant densities $u_1$ and $u_2$ on this mesh. Note that the linear correlation takes values $-1\le\rho(U_1,U_2)\le1$, and a larger correlation corresponds to a more in-phase-like appearance of both plant patterns. 

\begin{figure}
	\centering
	\subfloat[Numerical PDE simulations\label{fig: Multispeces pattern: correlation: D plots PDE sim}]{\includegraphics[width=\textwidth]{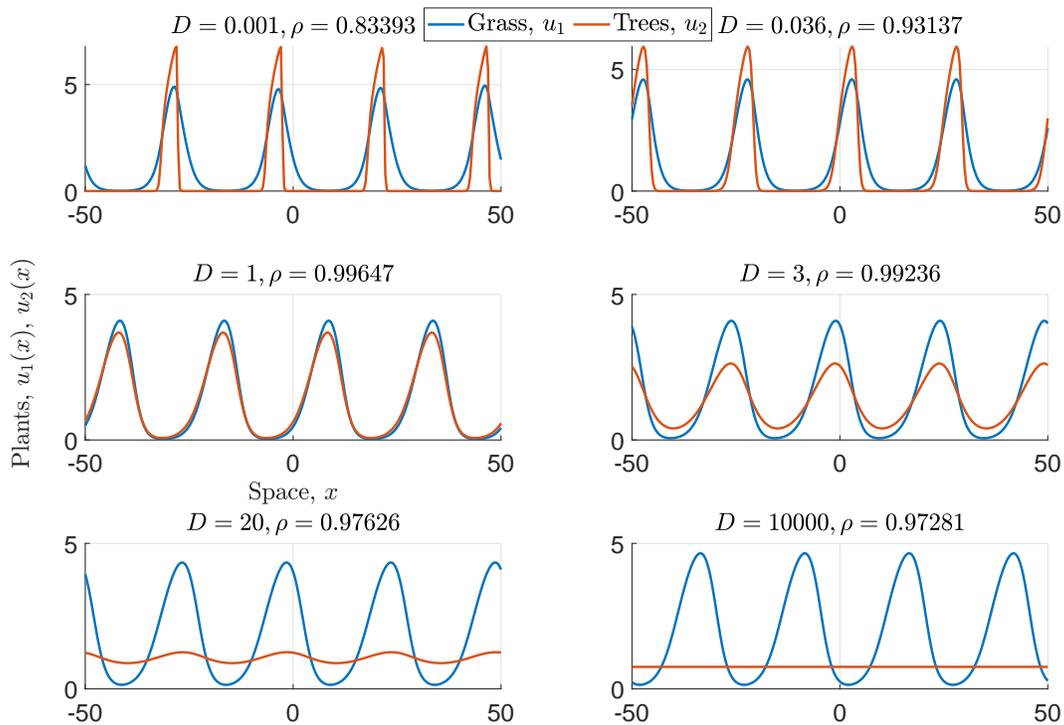}} \\
	\begin{minipage}{0.4\textwidth}
		\subfloat[Correlation between species\label{fig: Multispeces pattern: correlation: D plots correlation}]{\includegraphics[width=\textwidth]{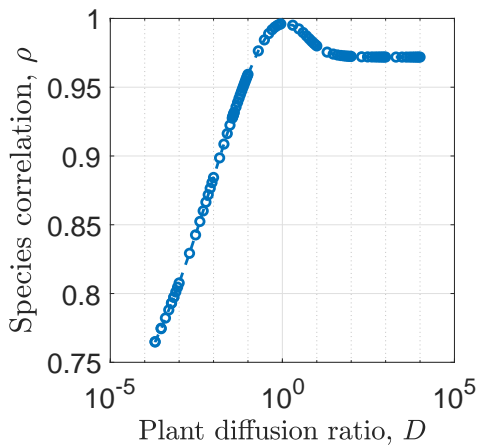}}
	\end{minipage}
	\hfill
	\begin{minipage}{0.58\textwidth}
		\caption{\textbf{Plant dispersal influences spatial species distribution and enables coexistence of a spatially uniform fast disperser with a patterned slow disperser.} The spatial correlation between plant species is shown in (b) and some example solutions are displayed in {\color{changes}(a)}. Note that the spatial correlation peaks close to $D=1$ {\color{changes}but does not reach unity due to the plant species differing in other parameters. No other parameters have any qualitative impact on correlation. In particular, species correlation is unaffected by changes in the strengths of local intraspecific competition, which are set to $k_1=k_2=10$ for visualisation purposes. For $D>1$}, coexistence of the locally superior species (which also disperses faster) in a spatially uniform state with a patterned state of the superior coloniser (but slower disperser) is possible. The species difference is set to $\chi=0.3$ and the wavelength $L$ is fixed to $L=25$ in the numerical continuation with the uphill migration speed allowed to vary.}\label{fig: Multispeces pattern: correlation: D plots}
	\end{minipage}	
\end{figure}

An exhaustive calculation of the linear correlation in the parameter space can be performed, as numerical continuation allows for an easy generation of model solutions. The ratio of the plant species' diffusion coefficients $D$ has the most significant impact on the correlation (Fig. \ref{fig: Multispeces pattern: correlation: D plots}). To specifically focus on the coexistence of grasses and trees, I have outlined in \cite{Eigentler2020coexistence_pattern} that if the species with slower growth also disperses at a slower rate (i.e. $(F-1)(D-1)>0$), then larger differences in the diffusion coefficients yield smaller spatial correlations, as the uphill edge of each vegetation band features a high density of the faster disperser only. In this parameter setting, that species can be referred to as the \textit{pioneer species}, as it is responsible for the colonisation of the bare ground in the uphill direction, before its competitor species utilises the increased resource availability in the newly colonised ground. {\color{changes}It is noteworthy that the species correlation of solutions of \eqref{eq: Multispecies pattern: Model: nondimensional model} is always positive. In particular, the plant densities never occur in antiphase, i.e. no complete spatial segregation of species takes place in the system.} Increases in the similarities of the species' dispersal behaviour causes an increase in the spatial correlation. In particular, the correlation attains its maximum value close to $D=1$, i.e. where both plant species diffuse at the same rate. {\color{changes}For $D=1$, the solution profile shows both plant species to be approximately in phase (Fig. \ref{fig: Multispeces pattern: correlation: D plots PDE sim}), but the influence of other parameters prevents the species from appearing exactly in phase. Nevertheless, changes to other parameters do not have any qualitative impact on species correlation in solutions of \eqref{eq: Multispecies pattern: Model: nondimensional model}.}

By contrast, if the assumption that one species both grows and disperses at a faster rate is dropped (i.e. if $(F-1)(D-1)<0$), then the correlation between the plant species does not decrease significantly from its maximum close to $D=1$ (Fig. \ref{fig: Multispeces pattern: correlation: D plots correlation}). However, the solution changes significantly. Instead of occurring in a patterned configuration with its competitor, the faster dispersing species attains a spatially uniform state, while the faster growing species (and slower disperser) remains in a patterned state (Fig. \ref{fig: Multispeces pattern: correlation: D plots PDE sim}). 

\section{Discussion}\label{sec: Multispecies pattern: discussion}

\begin{figure}
	\centering
	\includegraphics[width=\textwidth]{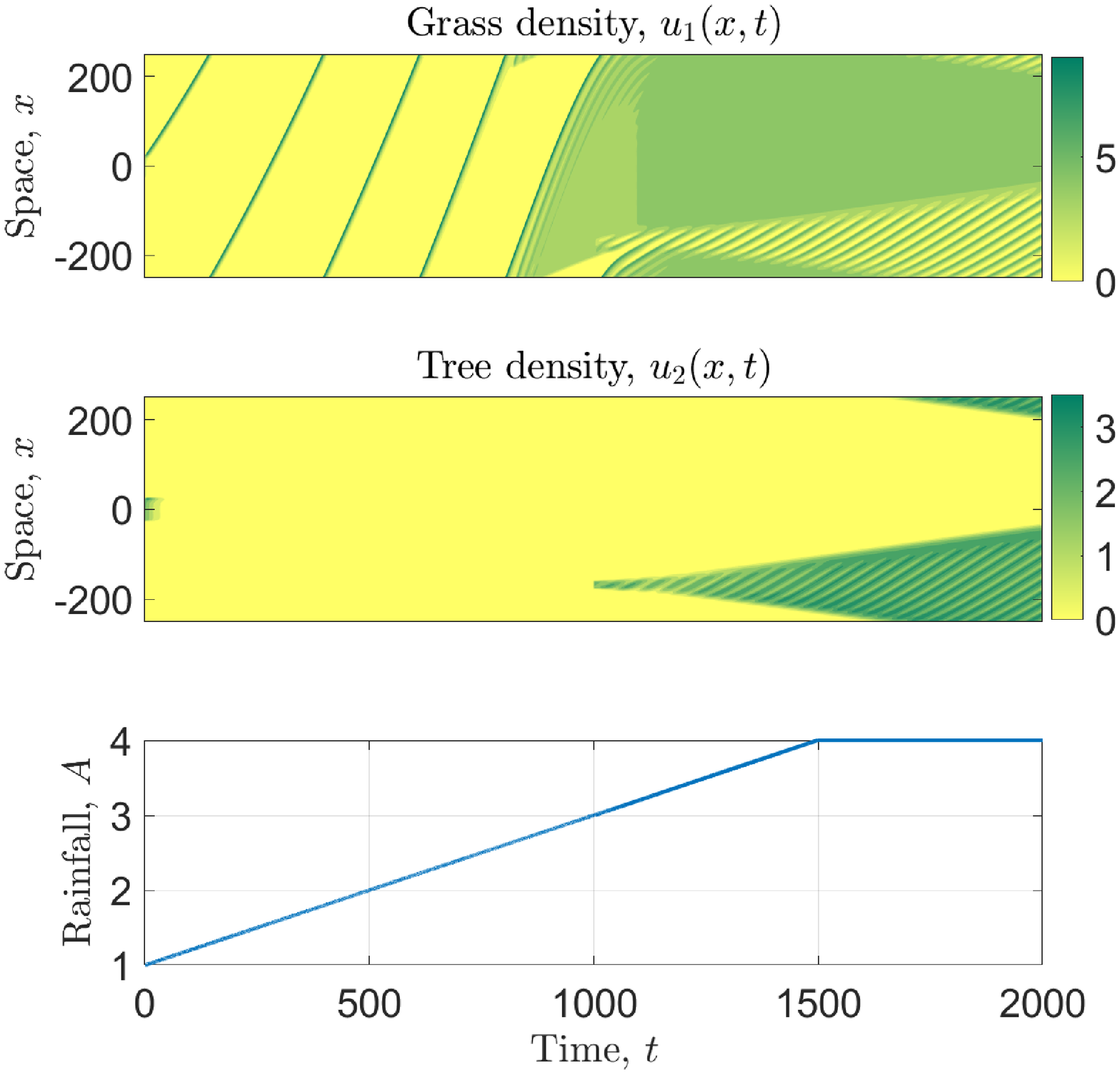}
	\caption{\textbf{Large species difference inhibits coexistence onset from desert.} Grass density $u_1$ and tree density $u_2$ of a model solution of \eqref{eq: Multispecies pattern: Model: nondimensional model} are shown in the $(t,x)$ under increasing precipitation volume $A$. Initially, both biomass densities are set to zero, apart from a region in the centre of the domain. The tree species becomes extinct and onset of a single-species grass pattern occurs. Onset of a coexistence pattern is only possible after a reintroduction of species $u_2$ at $t=1000$, following a sufficient increase in precipitation $A$. A further increase in $A$ causes a transition from the single-species grass pattern to a spatially uniform single-species state, but the coexistence pattern eventually invades. The parameter values are consistent with the bifurcation diagram shown in Fig. \ref{fig: Multispecies pattern: bifurcation diag chi 09}.}\label{fig: Multispecies pattern: discussion: onset from desert}
\end{figure}

The inclusion of {\color{changes}local} intraspecific competition dynamics in the modelling framework of the Klausmeier model for dryland vegetation patterns has a significant impact on the model solutions. In the context of the single-species model \eqref{eq: Multispecies pattern: single-species model}, {\color{changes}only considering intraspecific competition for water that acts on a long spatial scale} leads to an overestimation of the precipitation range in which patterns occur, while in the multispecies model \eqref{eq: Multispecies pattern: Model: nondimensional model}, {\color{changes}local} intraspecific competition is a key ingredient in the successful capture of species coexistence in a solution type that represents patterned vegetation.

In the single-species Klausmeier model, the rate of plant growth grows without bound as the plant density increases \cite{Klausmeier1999}. One possible motivation for this simplistic description is the type of ecosystem the modelling framework is describing. Dryland vegetation is limited by the low volumes of precipitation in arid ecosystems and thus total biomass is commonly low. Thus, {\color{changes}intraspecific competition among plants is generally only associated with long-range competition for water and} any negative density-dependent effects on the rate of plant growth caused by {\color{changes}local} intraspecific competition are neglected in the Klausmeier model and similar modelling frameworks \cite{Klausmeier1999, Rietkerk2002, HilleRisLambers2001}. However, even though total biomass on the ecosystem-wide scale is low, the spatial self-organisation of plants leads to the occurrence of localised patches in which biomass is high, thus raising a potential issue for the assumption to neglect {\color{changes}local} intraspecific competition.

Indeed, model solutions of the Klausmeier model and its extensions typically undergo several wavelength changes in their transition from a uniformly vegetated state to a desert state along the precipitation gradient. Towards the lower end of the rainfall range supporting stable patterns, the solutions' wavelength become large and biomass may locally increase to biologically unrealistic levels \cite{Bennett2019}. The consideration of {\color{changes}local} intraspecific competition dynamics in the single-species model \eqref{eq: Multispecies pattern: single-species model} presented in this paper does not allow for such solutions due to the existence of an upper bound, the maximum standing biomass, on the plant density at every space point. As a consequence, the patterned state loses stability (and existence) to the desert equilibrium at higher precipitation volumes than in the model without {\color{changes}local} intraspecific competition (Fig. \ref{fig: Multispecies pattern: single species: pattern existence and stability}). Hence, it can be concluded that models {\color{changes}that only consider intraspecific competition for water} overestimate the resilience of vegetation patterns to increasing aridity and that an understanding of intraspecific competition dynamics is essential to make predictions on desertification processes in ecosystems.

A characteristic feature of banded vegetation is the uphill migration of vegetation stripes \cite{Deblauwe2012}. Model solutions of the Klausmeier model consistently predict a reduction in uphill migration speed before a destabilisation due to increasing aridity occurs \cite{Bennett2019, Sherratt2013}, a property that can be used for early detection of degradation processes. While the introduction of {\color{changes}local} intraspecific competition to the single-species Klausmeier model decreases the size of the rainfall range supporting stable patterns, it stabilises patterned solutions with slower uphill migration speeds (Fig. \ref{fig: Multispecies pattern: single species: pattern existence and stability}). This further emphasises the importance of taking {\color{changes}local} intraspecific competition dynamics into account when developing methods of predicting future ecosystem developments, as they have a significant impact on ecologically important properties of model solutions.

The impact of {\color{changes}local} intraspecific competition in the framework of the multispecies model \eqref{eq: Multispecies pattern: Model: nondimensional model} is even more significant, because it stabilises species coexistence in both a spatially uniform state and in a state representing vegetation patterns (i.e. oscillations between a high level of biomass and zero). In the absence of {\color{changes}local} intraspecific competition dynamics, species coexistence can only occur in a spatially nonuniform savanna-type state (i.e. oscillations between two nonzero biomass levels) \cite{Eigentler2020savanna_coexistence}. {\color{changes}The main mechanism that enables coexistence in both \eqref{eq: Multispecies pattern: Model: nondimensional model} and the model neglecting {\color{changes}local} intraspecific competition is the spatial self-organisation of vegetation, which causes heterogeneities in the environmental conditions and thus gives rise to the existence of two behavioural niches (e.g. \cite{Whittaker1973}); that of colonisation and that of local superiority. In other words, coexistence is possible if the species which is locally inferior is superior in its colonisation abilities.} The latter allows the species to utilise the spatial heterogeneities in the resource availability to colonise new ground, before eventually being outcompeted locally by a second species \cite{Eigentler2020savanna_coexistence}. {\color{changes}With intraspecific competition dynamics restricted to competition for water}, such a balance is only maintained for relatively high volumes of precipitation, thus giving rise to the savanna-type model solution. As precipitation decreases, the coexistence state loses its stability to a single-species state of the coloniser species, as the beneficial effects of the coloniser's ability to self-organise itself into patterns tips the balance in its favour \cite{Eigentler2020savanna_coexistence}. If {\color{changes}local} intraspecific competition of the coloniser species is sufficiently strong, however, its advantages due to its self-organisation abilities decline as the maximum density in single plant patches declines. This stabilises the coexistence state at lower rainfall volumes at which it represents a vegetation pattern state (Fig. \ref{fig: Multispecies pattern: bifurcation diag ks separate}). This stabilisation of coexistence is related to classical results from nonspatial Lotka-Volterra competition models which state that coexistence is possible if intraspecific competition among all species is stronger than interspecific competition between them (e.g. \cite{Chesson2000}). The crucial difference is that due to the spatial self-organisation in the system, strong {\color{changes}local} intraspecific competition of one species only suffices to explain species coexistence \cite{Eigentler2020coexistence_pattern}.

Variations in the strength of {\color{changes}local} intraspecific competition of both species further have an impact on the system's bifurcation structure, and in particular on the onset of patterns. Decreases in {\color{changes}local} intraspecific competition strength cause a transition of the pattern onset mechanism at high precipitation levels from a Hopf bifurcation of the spatially uniform coexistence equilibrium to a stability change of a single-species pattern to the introduction of a second species (Fig. \ref{fig: Multispecies pattern: bifurcation diag k medium} and \ref{fig: Multispecies pattern: bifurcation diag k large}). As a consequence, model results predict that under weak {\color{changes}local} intraspecific competition no transition from a spatially uniform coexistence state to a patterned state can occur. Instead, one species' biomass decreases to zero as aridity increases, causing a transition to a spatially uniform single-species state. Only a reintroduction of the extinct species after a further decrease in precipitation can result in a patterned coexistence state.

The mechanism causing onset of coexistence patterns at the lower end of the precipitation range supporting their existence mainly depends on the difference between both species. If species are sufficiently similar, onset occurs at a homoclinic solution, while otherwise onset occurs due to a stability change of a single-species pattern to the introduction of a competitor (Fig. \ref{fig: Multispecies pattern: bifurcation diag chi}). This has significant ecological consequences as this predicts that the introduction of two significantly differing species into a desert state under sufficiently high precipitation volumes will not result in a successful invasion of the coexistence state. Instead, one species will become extinct and only a single-species pattern will prevail (Fig. \ref{fig: Multispecies pattern: discussion: onset from desert}). A transition to a coexistence state only becomes possible after a further increase in rainfall and a reintroduction of the second species. This, combined with the insights into ecosystem resilience presented above, highlights that mathematical modelling can be a powerful aid for the development of conservation programs in drylands.

The various hypotheses proposed by both \eqref{eq: Multispecies pattern: Model: nondimensional model} and \eqref{eq: Multispecies pattern: single-species model} could be tested using empirical data. However, the acquisition of data from vegetation patterns that are of sufficiently high quality and quantity is a significant challenge yet to be addressed by ecologists. Exceptions, for example on the uphill migration speed of vegetation stripes in various sites worldwide, exist \cite{Deblauwe2012} but in isolation such datasets are not sufficient to provide empirical tests for the models presented in this paper. Methods for data collection (in particular image processing) are expected to improve and thus such tests may become possible in the future.

{\color{changes}While the modelling framework presented in this paper leads to the successful capture of species coexistence in banded vegetation patterns, its counterpart neglecting local intraspecific competition dynamics only captures one out of many different types of savanna states \cite{Scholes1993}. Indeed, a comprehensive analysis of species correlation in coexistence solutions throughout the whole parameter space shows that both species' biomass densities are always approximately in phase (Fig. \ref{fig: Multispeces pattern: correlation: D plots PDE sim}). An exception occurs if the species with lower biomass yield per unit water consumed disperses significantly faster than its competitor. In this case, that species attains a spatially uniform solution but its competitor species remains in a spatial pattern. This is reminiscent of a different common savanna state: isolated clusters of trees within grasslands \cite{Sankaran2004}. However, under the assumptions taken in the modelling framework presented in this paper, my analysis predicts that such a state is only attained if woody species are superior in their water-to-biomass conversion abilities ($F>1$). Parameter estimates for dryland vegetation predict that grasses can convert water into new biomass more efficiently than trees or have a faster growth rate \cite{Klausmeier1999, Accatino2010} and I thus argue that the modelling framework presented by \eqref{eq: Multispecies pattern: Model: nondimensional model} is unable to capture such a type of savanna state. Instead, a potential mechanism causing this kind of coexistence is the competition for a second limiting resource (e.g. light). Competition for two resources can both prevent competitive exclusion (e.g. \cite{Chesson2000}) and cause multistability of single-species equilibria in mathematical frameworks. This can lead to the occurrence of localised patterns of one species within an otherwise uniform state of the second species, representing isolated clusters of trees within grasslands \cite{Kyriazopoulos2014}.

	The {\color{changes}local} intraspecific competition dynamics among plant species are incorporated into the modelling framework in a general way by combining them into one single variable, the maximum standing biomass, for each species. The significant impact of strong {\color{changes}local} intraspecific competition proposed by the results presented in this paper motivates a more detailed investigation of its details in the future. Promising first steps have been taken through the explicit modelling of toxic soil compounds produced by plants which inhibit their growth \cite{Marasco2019}. In the absence of water scarcity, these dynamics are sufficient to create a pattern-inducing feedback and give rise to yet another spatially patterned solution type typically referred to as a savanna state: spatial segregation of species, i.e. patterns that are antiphase. Even though this approach cannot make any statements about coexistence in water-deprived banded vegetation, it highlights the importance of local intraspecific competition dynamics. Moreover, it could be the foundation for a more detailed investigation of their impact on the competition and coexistence dynamics, potentially resulting in a modelling framework that unifies existing hypothesis on coexistence in vegetation patterns and savannas and thus allows for better predictions of future ecosystem dynamics. }

The modelling framework presented in this paper is very general and provides a deliberately simple description of a self-organisation principle in ecology. Moreover, results presented in this paper only depend on basic species properties but do not rely on any species-specific assumptions. This suggests that results may be extended to a host of different consumer-resource ecosystems in which coexistence of consumer species occurs. Indeed, the significant impact of self-organisation in such ecosystems has been addressed in recent years through both empirical and theoretical approaches \cite{Cornacchia2018, Christianen2017}, which emphasise that pattern formation can play a significant role in species coexistence and suggest more detailed theoretical studies of the phenomenon in the future to advance our understanding of species coexistence.

\section*{Acknowledgements}	
I thank Jonathan A Sherratt for helpful discussions and useful comments on the manuscript.

\section*{Funding information}
Lukas Eigentler was supported by The Maxwell Institute Graduate School in Analysis and its Applications, a Centre for Doctoral Training funded by the UK Engineering and Physical Sciences Research Council (grant EP/L016508/01), the Scottish Funding Council, Heriot-Watt University and the University of Edinburgh.

\clearpage
\bibliography{bibliography}

\begin{thebibliography}{10}
\expandafter\ifx\csname url\endcsname\relax
  \def\url#1{\texttt{#1}}\fi
\expandafter\ifx\csname doi\endcsname\relax
  \def\doi#1{\burlalt{doi:#1}{http://dx.doi.org/#1}}\fi
\expandafter\ifx\csname urlprefix\endcsname\relax\def\urlprefix{URL }\fi
\expandafter\ifx\csname href\endcsname\relax
  \def\href#1#2{#2}\fi
\expandafter\ifx\csname burlalt\endcsname\relax
  \def\burlalt#1#2{\href{#2}{#1}}\fi

\bibitem{Accatino2010}
F.~Accatino, C.~{De Michele}, R.~Vezzoli, D.~Donzelli, and R.~J. Scholes.
\newblock Tree{\textendash}grass co-existence in savanna: Interactions of rain
  and fire.
\newblock {\em J. Theor. Biol.}, 267(2):235--242, 2010.
\newblock \doi{10.1016/j.jtbi.2010.08.012}.

\bibitem{Baudena2013}
M.~Baudena and M.~Rietkerk.
\newblock Complexity and coexistence in a simple spatial model for arid savanna
  ecosystems.
\newblock {\em Theor. Ecol.}, 6(2):131--141, 2013.
\newblock \doi{10.1007/s12080-012-0165-1}.

\bibitem{Bennett2019}
J.~J.~R. Bennett and J.~A. Sherratt.
\newblock Long-distance seed dispersal affects the resilience of banded
  vegetation patterns in semi-deserts.
\newblock {\em J. Theor. Biol.}, 481:151--161, 2018.
\newblock \doi{10.1016/j.jtbi.2018.10.002}.

\bibitem{Champneys1996}
A.~R. Champneys, Y.~A. Kuznetsov, and B.~Sandstede.
\newblock A numerical toolbox for homoclinic bifurcation analysis.
\newblock {\em Int. J. Bifurcation Chaos}, 06(05):867--887, 1996.
\newblock \doi{10.1142/s0218127496000485}.

\bibitem{Chesson2000}
P.~Chesson.
\newblock Mechanisms of maintenance of species diversity.
\newblock {\em Annu. Rev. Ecol. Syst.}, 31:343--366, 2000.
\newblock \urlprefix\url{http://www.jstor.org/stable/221736}.

\bibitem{Christianen2017}
M.~Christianen, T.~{van der Heide}, S.~Holthuijsen, K.~{van der Reijden},
  A.~Borst, and H.~Olff.
\newblock Biodiversity and food web indicators of community recovery in
  intertidal shellfish reefs.
\newblock {\em Biol. Conserv.}, 213:317--324, 2017.
\newblock \doi{10.1016/j.biocon.2016.09.028}.

\bibitem{Consolo2019a}
G.~Consolo and G.~Valenti.
\newblock Secondary seed dispersal in the {K}lausmeier model of vegetation for
  sloped semi-arid environments.
\newblock {\em Ecol. Modell.}, 402:66--75, 2019.
\newblock \doi{10.1016/j.ecolmodel.2019.02.009}.

\bibitem{Cornacchia2018}
L.~Cornacchia, J.~{van de Koppel}, D.~{van der Wal}, G.~Wharton, S.~Puijalon,
  and T.~J. Bouma.
\newblock Landscapes of facilitation: how self-organized patchiness of aquatic
  macrophytes promotes diversity in streams.
\newblock {\em Ecology}, 99(4):832--847, 2018.
\newblock \doi{10.1002/ecy.2177}.

\bibitem{Deblauwe2012}
V.~Deblauwe, P.~Couteron, J.~Bogaert, and N.~Barbier.
\newblock Determinants and dynamics of banded vegetation pattern migration in
  arid climates.
\newblock {\em Ecol. Monogr.}, 82(1):3--21, 2012.
\newblock \doi{10.1890/11-0362.1}.

\bibitem{Dickovick2014}
J.~T. Dickovick.
\newblock {\em Africa 2014-2015}.
\newblock World Today (Stryker). Rowman \& Littlefield Publishers, 2014.

\bibitem{AUTO}
E.~J. Doedel, B.~E. Oldeman, A.~R. Champneys, F.~Dercole, T.~Fairgrieve,
  Y.~Kuznetsov, R.~Paenroth, B.~Sandstede, X.~Wang, and C.~Zhang.
\newblock Auto-07p: Continuation and bifurcation software for oridinary
  differential equations.
\newblock Technical report, 2012.

\bibitem{Eigentler2020coexistence_pattern}
L.~Eigentler.
\newblock Species coexistence in vegetation patterns facilitated by the
  interplay of spatial self-organisation and intraspecific competition.
\newblock {\em bioRxiv preprint}, 2020.
\newblock \doi{10.1101/2020.01.13.903179}.

\bibitem{Eigentler2019integrodifference}
L.~Eigentler and J.~A. Sherratt.
\newblock An integrodifference model for vegetation patterns in semi-arid
  environments with seasonality.
\newblock \burlalt{1911.10964}{http://arxiv.org/abs/1911.10964}.

\bibitem{Eigentler2018nonlocalKlausmeier}
L.~Eigentler and J.~A. Sherratt.
\newblock Analysis of a model for banded vegetation patterns in semi-arid
  environments with nonlocal dispersal.
\newblock {\em J. Math. Biol.}, 77(3):739--763, 2018.
\newblock \doi{10.1007/s00285-018-1233-y}.

\bibitem{Eigentler2019Multispecies}
L.~Eigentler and J.~A. Sherratt.
\newblock Metastability as a coexistence mechanism in a model for dryland
  vegetation patterns.
\newblock {\em Bull. Math. Biol.}, 81(7):2290--2322, 2019.
\newblock \doi{10.1007/s11538-019-00606-z}.

\bibitem{Eigentler2018impulsiveflat}
L.~Eigentler and J.~A. Sherratt.
\newblock Effects of precipitation intermittency on vegetation patterns in
  semi-arid landscapes.
\newblock {\em Physica D}, 405:132396, 2020.
\newblock \doi{10.1016/j.physd.2020.132396}.

\bibitem{Eigentler2020savanna_coexistence}
L.~Eigentler and J.~A. Sherratt.
\newblock Spatial self-organisation enables species coexistence in a model for
  savanna ecosystems.
\newblock {\em J. Theor. Biol.}, 487:110122, 2020.
\newblock \doi{10.1016/j.jtbi.2019.110122}.

\bibitem{Gandhi2018}
P.~Gandhi, L.~Werner, S.~Iams, K.~Gowda, and M.~Silber.
\newblock A topographic mechanism for arcing of dryland vegetation bands.
\newblock {\em Journal of The Royal Society Interface}, 15(147):20180508, 2018.
\newblock \doi{10.1098/rsif.2018.0508}.

\bibitem{Gilad2007a}
E.~Gilad, M.~Shachak, and E.~Meron.
\newblock Dynamics and spatial organization of plant communities in
  water-limited systems.
\newblock {\em Theor. Popul. Biol.}, 72(2):214--230, 2007.
\newblock \doi{10.1016/j.tpb.2007.05.002}.

\bibitem{Gilad2004}
E.~Gilad, J.~von Hardenberg, A.~Provenzale, M.~Shachak, and E.~Meron.
\newblock Ecosystem engineers: From pattern formation to habitat creation.
\newblock {\em Phys. Rev. Lett.}, 93:098105, 2004.
\newblock \doi{10.1103/PhysRevLett.93.098105}.

\bibitem{Gilad2007}
E.~Gilad, J.~von Hardenberg, A.~Provenzale, M.~Shachak, and E.~Meron.
\newblock A mathematical model of plants as ecosystem engineers.
\newblock {\em J. Theor. Biol.}, 244(4):680 -- 691, 2007.
\newblock \doi{j.jtbi.2006.08.006}.

\bibitem{Gowda2018}
K.~Gowda, S.~Iams, and M.~Silber.
\newblock Signatures of human impact on self-organized vegetation in the {H}orn
  of {A}frica.
\newblock {\em Sci. Rep.}, 8:1--8, 2018.
\newblock \doi{10.1038/s41598-018-22075-5}.

\bibitem{HilleRisLambers2001}
R.~HilleRisLambers, M.~Rietkerk, F.~{van den Bosch}, H.~H.~T. Prins, and H.~{de
  Kroon}.
\newblock Vegetation pattern formation in semi-arid grazing systems.
\newblock {\em Ecology}, 82(1):50--61, 2001.
\newblock \doi{10.2307/2680085}.

\bibitem{Klausmeier1999}
C.~A. Klausmeier.
\newblock Regular and irregular patterns in semiarid vegetation.
\newblock {\em Science}, 284(5421):1826--1828, 1999.
\newblock \doi{10.1126/science.284.5421.1826}.

\bibitem{Kyriazopoulos2014}
P.~Kyriazopoulos, J.~Nathan, and E.~Meron.
\newblock Species coexistence by front pinning.
\newblock {\em Ecol. Complexity}, 20:271--281, 2014.
\newblock \doi{10.1016/j.ecocom.2014.05.001}.

\bibitem{Lefever1997}
R.~Lefever and O.~Lejeune.
\newblock On the origin of tiger bush.
\newblock {\em Bull. Math. Biol.}, 59(2):263--294, 1997.
\newblock \doi{10.1007/bf02462004}.

\bibitem{Marasco2019}
A.~Marasco, F.~Giannino, and A.~Iuorio.
\newblock Modelling competitive interactions and plant{\textendash}soil
  feedback in vegetation dynamics.
\newblock {\em Ricerche di Matematica}, mar 2020.
\newblock \doi{10.1007/s11587-020-00497-6}.

\bibitem{Marasco2014}
A.~Marasco, A.~Iuorio, F.~Carteni, G.~Bonanomi, D.~M. Tartakovsky,
  S.~Mazzoleni, and F.~Giannino.
\newblock Vegetation pattern formation due to interactions between water
  availability and toxicity in plant{\textendash}soil feedback.
\newblock {\em Bull. Math. Biol.}, 76(11):2866--2883, 2014.
\newblock \doi{10.1007/s11538-014-0036-6}.

\bibitem{Martinez-Garcia2013a}
R.~Martinez-Garcia, J.~M. Calabrese, E.~Hernandez-Garcia, and C.~Lopez.
\newblock Vegetation pattern formation in semiarid systems without facilitative
  mechanisms.
\newblock {\em Geophysical Research Letters}, 40(23):6143--6147, dec 2013.
\newblock \doi{10.1002/2013gl058797}.

\bibitem{Martinez-Garcia2014}
R.~Martinez-Garcia, J.~M. Calabrese, E.~Hernandez-Garcia, and C.~Lopez.
\newblock Minimal mechanisms for vegetation patterns in semiarid regions.
\newblock {\em Philosophical Transactions of the Royal Society A: Mathematical,
  Physical and Engineering Sciences}, 372(2027):20140068, oct 2014.
\newblock \doi{10.1098/rsta.2014.0068}.

\bibitem{Martinez-Garcia2018}
R.~Martinez-Garcia and C.~Lopez.
\newblock From scale-dependent feedbacks to long-range competition alone: a
  short review on pattern-forming mechanisms in arid ecosystems.
\newblock 2018, \burlalt{1801.01399v1}{http://arxiv.org/abs/1801.01399v1}.

\bibitem{Mazzoleni2015}
S.~Mazzoleni, G.~Bonanomi, G.~Incerti, M.~L. Chiusano, P.~Termolino, A.~Mingo,
  M.~Senatore, F.~Giannino, F.~Carteni, M.~Rietkerk, and V.~Lanzotti.
\newblock Inhibitory and toxic effects of extracellular self-{DNA} in litter: a
  mechanism for negative plant-soil feedbacks?
\newblock {\em New Phytol.}, 205(3):1195--1210, 2015.
\newblock \doi{10.1111/nph.13121}.

\bibitem{Meron2016}
E.~Meron.
\newblock Pattern formation - a missing link in the study of ecosystem response
  to environmental changes.
\newblock {\em Math. Biosci.}, 271:1--18, 2016.
\newblock \doi{10.1016/j.mbs.2015.10.015}.

\bibitem{Meron2019}
E.~Meron, J.~J.~R. Bennett, C.~Fernandez-Oto, O.~Tzuk, Y.~R. Zelnik, and
  G.~Grafi.
\newblock Continuum modeling of discrete plant communities: Why does it work
  and why is it advantageous?
\newblock {\em Mathematics}, 7(10):987, 2019.
\newblock \doi{10.3390/math7100987}.

\bibitem{Nathan2013}
J.~Nathan, J.~{von Hardenberg}, and E.~Meron.
\newblock Spatial instabilities untie the exclusion-principle constraint on
  species coexistence.
\newblock {\em J. Theor. Biol.}, 335:198--204, 2013.
\newblock \doi{10.1016/j.jtbi.2013.06.026}.

\bibitem{Rademacher2007}
J.~D. Rademacher, B.~Sandstede, and A.~Scheel.
\newblock Computing absolute and essential spectra using continuation.
\newblock {\em Physica D}, 229(2):166--183, 2007.
\newblock \doi{10.1016/j.physd.2007.03.016}.

\bibitem{Reynolds2007}
J.~F. Reynolds, D.~M.~S. Smith, E.~F. Lambin, B.~L. Turner, M.~Mortimore,
  S.~P.~J. Batterbury, T.~E. Downing, H.~Dowlatabadi, R.~J. Fernandez, J.~E.
  Herrick, E.~Huber-Sannwald, H.~Jiang, R.~Leemans, T.~Lynam, F.~T. Maestre,
  M.~Ayarza, and B.~Walker.
\newblock Global desertification: Building a science for dryland development.
\newblock {\em Science}, 316(5826):847--851, 2007.
\newblock \doi{10.1126/science.1131634}.

\bibitem{Rietkerk2002}
M.~Rietkerk, M.~C. Boerlijst, F.~{van Langevelde}, R.~HilleRisLambers, J.~{van
  de Koppel}, L.~Kumar, H.~H.~T. Prins, and A.~M. {de Roos}.
\newblock Self‐organization of vegetation in arid ecosystems.
\newblock {\em Am. Nat.}, 160(4):524--530, 2002.
\newblock \doi{10.1086/342078}.

\bibitem{Rietkerk2008}
M.~Rietkerk and J.~{van de Koppel}.
\newblock Regular pattern formation in real ecosystems.
\newblock {\em Trends Ecol. Evol.}, 23(3):169 -- 175, 2008.
\newblock \doi{10.1016/j.tree.2007.10.013}.

\bibitem{Saco2018}
P.~M. Saco, M.~{Moreno-de las Heras}, S.~Keesstra, J.~Baartman, O.~Yetemen, and
  J.~F. Rodriguez.
\newblock Vegetation and soil degradation in drylands: Non linear feedbacks and
  early warning signals.
\newblock {\em Curr. Opin. Environ. Sci. Health}, 5:67--72, 2018.
\newblock \doi{10.1016/j.coesh.2018.06.001}.

\bibitem{Sankaran2004}
M.~Sankaran, J.~Ratnam, and N.~P. Hanan.
\newblock Tree-grass coexistence in savannas revisited - insights from an
  examination of assumptions and mechanisms invoked in existing models.
\newblock {\em Ecology Letters}, 7(6):480--490, jun 2004.
\newblock \doi{10.1111/j.1461-0248.2004.00596.x}.

\bibitem{Scholes1993}
R.~J. Scholes and B.~H. Walker.
\newblock {\em An African Savanna}.
\newblock Cambridge University Press, 1993.
\newblock \doi{10.1017/cbo9780511565472}.

\bibitem{Seghieri1997}
J.~Seghieri, S.~Galle, J.~Rajot, and M.~Ehrmann.
\newblock Relationships between soil moisture and growth of herbaceous plants
  in a natural vegetation mosaic in {N}iger.
\newblock {\em J. Arid. Environ.}, 36(1):87--102, 1997.
\newblock \doi{10.1006/jare.1996.0195}.

\bibitem{Sherratt2005}
J.~A. Sherratt.
\newblock An analysis of vegetation stripe formation in semi-arid landscapes.
\newblock {\em J. Math. Biol.}, 51(2):183--197, 2005.
\newblock \doi{10.1007/s00285-005-0319-5}.

\bibitem{Sherratt2010}
J.~A. Sherratt.
\newblock Pattern solutions of the {K}lausmeier model for banded vegetation in
  semi-arid environments {I}.
\newblock {\em Nonlinearity}, 23(10):2657--2675, 2010.
\newblock \doi{10.1088/0951-7715/23/10/016}.

\bibitem{Sherratt2011}
J.~A. Sherratt.
\newblock Pattern solutions of the {K}lausmeier model for banded vegetation in
  semi-arid environments {II}: patterns with the largest possible propagation
  speeds.
\newblock {\em Proc. R. Soc. Lond. A}, 467(2135):3272--3294, 2011.
\newblock \doi{10.1098/rspa.2011.0194}.

\bibitem{Sherratt2012}
J.~A. Sherratt.
\newblock Numerical continuation methods for studying periodic travelling wave
  (wavetrain) solutions of partial differential equations.
\newblock {\em Appl. Math. Comput.}, 218(9):4684--4694, 2012.
\newblock \doi{10.1016/j.amc.2011.11.005}.

\bibitem{Sherratt2013}
J.~A. Sherratt.
\newblock History-dependent patterns of whole ecosystems.
\newblock {\em Ecol. Complexity}, 14:8--20, 2013.
\newblock \doi{10.1016/j.ecocom.2012.12.002}.

\bibitem{Sherratt2013a}
J.~A. Sherratt.
\newblock Numerical continuation of boundaries in parameter space between
  stable and unstable periodic travelling wave (wavetrain) solutions of partial
  differential equations.
\newblock {\em Adv. Comput. Math.}, 39(1):175--192, 2013.
\newblock \doi{10.1007/s10444-012-9273-0}.

\bibitem{Sherratt2013III}
J.~A. Sherratt.
\newblock Pattern solutions of the {K}lausmeier model for banded vegetation in
  semi-arid environments {III}: The transition between homoclinic solutions.
\newblock {\em Physica D}, 242(1):30 -- 41, 2013.
\newblock \doi{10.1016/j.physd.2012.08.014}.

\bibitem{Sherratt2013IV}
J.~A. Sherratt.
\newblock Pattern solutions of the {K}lausmeier model for banded vegetation in
  semiarid environments {IV}: Slowly moving patterns and their stability.
\newblock {\em SIAM J. Appl. Math.}, 73(1):330--350, 2013.
\newblock \doi{10.1137/120862648}.

\bibitem{Sherratt2013V}
J.~A. Sherratt.
\newblock Pattern solutions of the {K}lausmeier model for banded vegetation in
  semiarid environments {V}: The transition from patterns to desert.
\newblock {\em SIAM J. Appl. Math.}, 73(4):1347--1367, 2013.
\newblock \doi{10.1137/120899510}.

\bibitem{Siero2018}
E.~Siero.
\newblock Nonlocal grazing in patterned ecosystems.
\newblock {\em J. Theor. Biol.}, 436:64--71, 2018.
\newblock \doi{10.1016/j.jtbi.2017.10.001}.

\bibitem{Siteur2014}
K.~Siteur, E.~Siero, M.~B. Eppinga, J.~D. Rademacher, A.~Doelman, and
  M.~Rietkerk.
\newblock Beyond {T}uring: The response of patterned ecosystems to
  environmental change.
\newblock {\em Ecol. Complexity}, 20:81 -- 96, 2014.
\newblock \doi{10.1016/j.ecocom.2014.09.002}.

\bibitem{Taft1997}
J.~B. Taft.
\newblock Savanna and open-woodland communities.
\newblock In {\em Conservation in Highly Fragmented Landscapes}, pages 24--54.
  Springer New York, 1997.
\newblock \doi{10.1007/978-1-4757-0656-7_2}.

\bibitem{Valentin1999}
C.~Valentin, J.~d'Herbès, and J.~Poesen.
\newblock Soil and water components of banded vegetation patterns.
\newblock {\em CATENA}, 37(1–2):1--24, 1999.
\newblock \doi{10.1016/S0341-8162(99)00053-3}.

\bibitem{Whittaker1973}
R.~H. Whittaker, S.~A. Levin, and R.~B. Root.
\newblock Niche, habitat, and ecotope.
\newblock {\em The American Naturalist}, 107(955):321--338, may 1973.
\newblock \doi{10.1086/282837}.

\bibitem{Zelnik2013}
Y.~R. Zelnik, S.~Kinast, H.~Yizhaq, G.~Bel, and E.~Meron.
\newblock Regime shifts in models of dryland vegetation.
\newblock {\em Philos. Trans. R. Soc. London, Ser. A}, 371(2004):20120358,
  2013.
\newblock \doi{10.1098/rsta.2012.0358}.

\end{thebibliography}

\end{document}